\documentclass[useAMS]{mn2e}
\pdfoutput=1
\usepackage{epsfig}
\usepackage{amsmath}
\usepackage{amssymb}

\def\be{\begin{equation}}
\def\ee{\end{equation}}
\def\ba{\begin{eqnarray}}
\def\ea{\end{eqnarray}}
\def\go{\mathrel{\raise.3ex\hbox{$>$}\mkern-14mu
             \lower0.6ex\hbox{$\sim$}}}
\def\lo{\mathrel{\raise.3ex\hbox{$<$}\mkern-14mu
             \lower0.6ex\hbox{$\sim$}}}
\def\bxi{{\mbox{\boldmath $\xi$}}}
\def\br{{\bf r}}

\def\orb{{\rm orb}}

\def\bC{{\bf C}}

\def\bOmega{{\bf \Omega}}

\def\omi{\omega_\alpha}

\begin{document}

\title[Dynamical Tides in KOI-54]{Dynamical Tides in Eccentric Binaries and Tidally-Excited Stellar Pulsations
in {\it KEPLER} KOI-54}
\author[J. Fuller and D. Lai]
{Jim Fuller\thanks{Email:
derg@astro.cornell.edu; dong@astro.cornell.edu}
and Dong Lai\\
Center for Space Research, Department of Astronomy, Cornell University, Ithaca, NY 14853, USA}

\label{firstpage}
\maketitle
\begin{abstract}

Recent observation of the tidally-excited stellar oscillations in the
main-sequence binary KOI-54 by the {\it KEPLER} satellite
provides a unique opportunity for studying dynamical tides in
eccentric binary systems.  We develop a general theory of tidal excitation
of oscillation modes of rotating binary stars, and apply our theory to
tidally excited gravity modes (g-modes) in KOI-54.  The strongest observed
oscillations, which occur at 90 and 91 times the orbital frequency,
are likely due to prograde $m=2$ modes (relative to the stellar spin
axis) locked in resonance with the orbit.  The remaining flux
oscillations with frequencies that are integer multiples of the
orbital frequency are likely due to nearly resonant $m=0$ g-modes;
such axisymmetric modes generate larger flux variations compared to
the $m=2$ modes, assuming that the spin inclination angle of the star
is comparable to the orbital inclination angle.  We examine the
process of resonance mode locking under the combined effects of dynamical
tides on the stellar spin and orbit and the intrinsic stellar
spindown. We show that KOI-54 can naturally evolve into a state in
which at least one $m=2$ mode is locked in resonance with the orbital
frequency. Our analysis provides an explanation for the fact that
only oscillations with frequencies less than 90-100 times the orbital
frequency are observed. We have also found evidence from the published
{\it KEPLER} result that three-mode nonlinear coupling occurs in the
KOI-54 system. We suggest that such nonlinear mode coupling may
explain the observed oscillations that are not harmonics of the
orbital frequency.

\end{abstract}


\section{Introduction}

Tides play an important role in binary star systems and in star-planet
systems. While numerous studies of tidal effects have been based on
the so-called equilibrium tide theory, which parametrizes tidal
dissipation by an effective tidal lag angle/time or tidal quality
factor (e.g., Darwin 1880; Goldreich \& Soter 1966; Alexander 1973; Hut 1981),
the underlying physics of tides in fluid stars and planets
involves dynamical excitations of waves and oscillations by the 
tidal force (see Ogilvie \& Lin 2007 and Zahn 2008 for recent reviews).
Tides in highly eccentric systems are particularly rich in their dynamical
behavior, since wave modes with a wide range of frequencies
can be excited and participate in the tidal interaction.
Various aspects of dynamical tides in eccentric binaries have been studied
by Lai (1996,1997), Kumar \& Quataert (1998), Witte \& Savonije (1999,2001),
Willems et al.~(2003), Ivanov \& Papaloizou (2004) and Papaloizou 
\& Ivanov (2010).

Recent observations of the binary star system HD 187091 
(KOI-54) by the {\it Kepler}
satellite provide a unique opportunity for studying dynamical tides in
eccentric binaries. KOI-54 consists of two A stars
(mass $M_{1,2}=2.32,\,2.38\,M_\odot$ and radius
$R_{1,2}=2.19,\,2.33\,R_\odot$) in an eccentric ($e=0.8342$) orbit with
period $P=41.805$~days (Welsh et al.~2011). The binary is nearly
face-on with orbital inclination $i_{\rm orb}=5.52^\circ$. In addition to
periodic brightening events caused by tidal distortion and irradiation
of the two stars during their close periastron passages, the power
spectrum of the {\it Kepler} light curve revealed 30 significant (with
a signal-to-noise ratio $\go 7$) stellar pulsation modes.  The
observed mode periods range from 45~hours to 11 hours, corresponding
to the mode frequency $f_\alpha$ ranging from $22.42\,f_\orb$ to
$91\,f_\orb$ (where $f_\orb=P^{-1}$ is the orbital frequency). Most
interestingly, twenty-one of these mode frequencies are integer
multiples of $f_\orb$ (with the ratio $f_\alpha/f_\orb$ differing from
an integer by $0.01$ or less). The two dominant modes have frequencies
that are exactly 90 and 91 times $f_\orb$, with the corresponding flux
amplitudes of 297.7~$\mu$mag and 229.4~$\mu$mag, respectively.

While dynamical tides in massive-star binaries have
been studied before (e.g., Zahn 1977, Goldreich \& Nicholson 1989 for
circular binaries; Lai 1996,1997, Kumar \& Quataert 1998 and Witte \&
Savonije 1999,2011 for eccentric binaries), KOI-54 represents the
first example where tidally excited oscillations are directly observed
and therefore serves as an explict demonstration of dynamical tides at
work in the system. As discussed in Welsh et al.~(2011), the observed
oscillation modes are puzzling: over 20 of the observed modes are
nearly exact integer multiples of the orbital frequency, yet several
others are not. It is not clear why the dominant modes are so
prominent, e.g., why modes with frequencies 90, 91, 44, and 40 times 
$f_{\rm orb}$ are clearly visible, and yet modes with frequencies 
greater than $91f_{\rm orb}$ and those less than $20f_{\rm orb}$
appear to be absent. 

The goal of this paper is to explain some of the observational
puzzles related to KOI-54 and to develop the general theoretical
framework for studying tidally-excited oscillations in eccentric
binary systems.

Our paper is organized as follows. In Section 2, we derive the
general equations for calculating the energies of tidally excited
oscillation modes in an eccentric binary. Our theory improves upon
previous (and less rigorous) works, and provides a clear relationship
between the resonant mode energy and non-resonant mode energy.  In
Section 3 we study the properties of non-radial g-modes relevant to 
the stars in the KOI-54 system and calculate the non-resonant mode
energies -- these serve as a benchmark for examining the effect of
resonances.  In Section 4 we present our calculations of the flux
variation due to tidally-forced oscillations. We show that the
observed flux variation in KOI-54 can be largely explained when a
high-frequency mode is locked into resonance (with the mode
frequency equal to $90f_{\rm orb}$). In Section 5 we study the
possibility of resonance locking. We show that the combination of the
secular tidal orbital/spin evolution and the intrinsic spindown of the
star (e.g., due to stellar evolution) may naturally lead to resonance
locking of a particular mode. Our analysis demonstrates that (in the KOI-54 system) a mode with frequency around
$90f_{\rm orb}$ can be resonantly locked, while modes with higher 
frequencies cannot.  In Section 6 we discuss the origin of the observed
modes in KOI-54 with frequencies that are not an integer multiple of 
$f_{\rm orb}$, including the evidence of nonlinear mode coupling.
We conclude in Section 7 with a discussion of future prospects and remaining puzzles.

\section{Dynamical Tide in Eccentric Binary Stars: General Theory}
\label{general theory}

We consider the tidally-excited oscillations of the primary star of
mass $M$ and radius $R$ by the companion of mass $M'$.  The
gravitational potential produced by $M'$ can be written as 
\be
U({\bf r}_i,t)=-GM'\sum_{lm}{W_{lm}r^l\over D^{l+1}}\,\, e^{-im\Phi(t)}
Y_{lm}(\theta,\phi_i), 
\ee 
where ${\bf r}_i=(r,\theta,\phi_i=\phi+\Omega_s t)$ 
is the position vector (in spherical
coordinates) relative to the center of star $M$
(the azimuthal angle $\phi$ is measured in the rotating frame of the star,
with the rotation rate $\Omega_s$ and the spin axis aligned with the
orbital angular momentum), 
$D(t)$ is the binary separation and 
$\Phi$ is the orbital true anomaly.
The dominant terms have $l=|m|=2$ and $l=2$, $m=0$, and for these terms
$W_{2\pm 2}=(3\pi/10)^{1/2}$ and $W_{20}=(\pi/5)^{1/2}$.  The linear response of
star $M$ is specified by the Lagrangian displacement $\bxi(\br,t)$,
which satisfies the equation of motion (in the rotating frame of the
star) 
\be
\frac{\partial^2 \bxi}{\partial t^2}+2\bOmega_s\times
\frac{\partial\bxi}{\partial t}+{\bC}\cdot\bxi=-\nabla U,
\label{eq:eqnmotion2}
\ee
where $\bC$ is a self-adjoint operator (a function of the pressure and
gravity perturbations) acting on $\bxi$ (see, e.g., 
Friedman \& Schutz 1978).
A free mode of frequency $\omega_\alpha$ (in the rotating frame) 
with $\bxi_\alpha(\br,t)=\bxi_\alpha(\br)\,e^{-i\omega_\alpha t}\propto
e^{im\phi-i\omega_\alpha t}$ satisfies 
\be
-\omi^2\bxi_\alpha-2i\omi\bOmega_s\times\bxi_\alpha+\bC\cdot
\bxi_\alpha=0, \ee 
where $\{\alpha\}$ denotes the mode index, which
includes the radial mode number $n$, the polar mode number $L$ 
(which reduces to $l$ for spherical stars) and the 
azimuthal mode number $m$. 
We carry out phase space mode expansion (Schenk et al.~2002)
\be
\left[\begin{array}{c}
\bxi\\
{\partial\bxi/\partial t}
\end{array}\right]
=\sum_\alpha c_\alpha(t)
\left[\begin{array}{c}
\bxi_\alpha(\br)\\
-i\omi\bxi_\alpha(\br)
\end{array}\right],
\ee
where the sum includes not only mode indices, but also
both positive and negative $\omega_\alpha$. Note that the
usual mode decomposition, $\bxi=\sum_{\alpha'} c_\alpha \bxi_\alpha$
(with the sum including only mode indices),
adopted in many previous studies (e.g., Lai 1997, Kumar \& Quataert 1998;
Witte \& Savonije 1999), are rigorously valid only for non-rotating stars.
Using the orthogonality relation
$\langle\bxi_\alpha,2i\bOmega_s\times\bxi_{\alpha'}\rangle+
(\omega_\alpha+\omega_{\alpha'})\langle\bxi_\alpha,\bxi_{\alpha'}
\rangle=0$ (for $\alpha\neq \alpha'$), where
$\langle A,B\rangle\equiv\int\!d^3x\,\rho\, (A^\ast\cdot B)$,
we find (Lai \& Wu 2006)\footnote{As noted before, in this paper
we restrict to aligned spin-orbit configurations for simplicity.
Generalization to misaligned systems is straightforward (Lai \& Wu 2006;
see also Ho \& Lai 1999).}
\begin{align}
{\dot c}_\alpha+(i\omi +\gamma_\alpha) c_\alpha &= {i\over 2\varepsilon_\alpha}\langle\bxi_\alpha(\br),-\nabla U
\rangle \nonumber \\
&= \frac{iGM'W_{lm}Q_\alpha}{2 \varepsilon_\alpha D^{l+1}} e^{im\Omega_s t-im\Phi},
\label{eq:adot}
\end{align}
where $\gamma_\alpha$ is the mode (amplitude) damping rate, and 
\begin{align}
&Q_{\alpha}\equiv\bigl\langle\bxi_\alpha,\nabla (r^lY_{lm})
\bigr\rangle,
\label{eq:Qdefine}\\
&\varepsilon_\alpha\equiv
\omi+\langle\bxi_\alpha,i\bOmega_s\times\bxi_{\alpha}\rangle,
\end{align}
and we have used the normalization $\langle\bxi_\alpha,\bxi_\alpha\rangle =1$. 
The quantity $Q_\alpha$ (called the ``tidal overlap integral'' or ``tidal
coupling coefficient'') directly relates to the tidally excited mode
amplitude. We shall focus on $l=2$, $m=0$ and $|m|=2$ modes in the following (although we will continue to
use the notations $l$ and $m$ so that it would be easy to generalize to 
high-order tides).

The general solution equation (\ref{eq:adot}) is 
\be
c_\alpha(t)=e^{-i\omega_\alpha t-\gamma_\alpha t}
\int_{t_0}^t \frac{iGM'W_{lm}Q_\alpha}{2 \varepsilon_\alpha D^{l+1}} e^{i\sigma_\alpha t+\gamma_\alpha t-im\Phi}\,dt,
\ee
assuming $c_\alpha(t_0)=0$, where 
\be
\sigma_\alpha=\omega_\alpha+ m\Omega_s
\ee
is the mode frequency in the inertial frame. Let $t_j=(2j-1)P/2$ 
(with $j=0,1,2,\cdots$) be the times at apastron. After the $k$th 
periastron passage, the mode amplitude becomes
\begin{align}
c_\alpha(t_k)=&(\Delta c_\alpha)\,e^{im\Omega_s t_k-
(i\sigma_\alpha+\gamma_\alpha)P/2}\nonumber \\
&\times\left[ {1-e^{-(i\sigma_\alpha+\gamma_\alpha)kP} \over 1-e^{-(i\sigma_\alpha+\gamma_\alpha)P}}\right],
\end{align}
with
\be
\label{deltacalpha}
\Delta c_\alpha=\int_{-P/2}^{P/2}\!dt\,
\frac{iGM'W_{lm}Q_\alpha}{2 \varepsilon_\alpha D^{l+1}} e^{i\sigma_\alpha+\gamma_\alpha t-im\Phi}.
\ee
For $\gamma_\alpha kP\gg 1$, the steady-state mode energy in the inertial frame 
\be
\label{Ealpha}
E_\alpha=2\sigma_\alpha\varepsilon_\alpha |c_\alpha|^2
\ee
becomes (Lai 1997) \footnote{Equation (\ref{eq:Ealpha}) was derived in 
Lai (1997) in an approximate manner (since mode decomposition was not
done rigorously), and physical arguments were used to get rid of
a fictitious term.}
\begin{align}
\label{eq:Ealpha}
E_\alpha &= {\Delta E_\alpha\over 2(\cosh\gamma_\alpha P-\cos\sigma_\alpha P)} \nonumber \\
&\simeq {\Delta E_\alpha\over 4\sin^2(\sigma_\alpha P/2)+(\gamma_\alpha P)^2},
\end{align}
where the second equality assumes $\gamma_\alpha P\ll 1$.
Here $\Delta E_\alpha$ is the energy transfer to the mode in the ``first''
periastron passage:
\be
\Delta E_\alpha={G{M'}^2\over R}\!\left({R\over D_p}\right)^{2(l+1)}
\!{2\pi^2\sigma_\alpha\over
\varepsilon_\alpha}\left|Q_\alpha K_{lm}(\sigma_\alpha)\right|^2,
\label{eq:dEalpha}
\ee
where $D_p=a(1-e)$ is the periastron distance ($a$ is the orbital 
semi-major axis) and
\be
\label{K}
K_{lm}(\sigma_\alpha)={W_{lm}\over 2\pi}\int_{-P/2}^{P/2}\!dt\,
\left({R\over D}\right)^{l+1}\! e^{i\sigma_\alpha t-im\Phi}.
\ee
Note that in equation (\ref{eq:dEalpha}), both $Q_\alpha$ and $K_{lm}$ 
are dimensionless (in units such that $G=M=R=1$).

Equation (\ref{eq:Ealpha}) shows that 
when $\sigma_\alpha P$ is not close to $2\pi N$ (where $N$ is an integer),
the steady-state mode energy is approximately $\Delta E_\alpha$. Thus
$\Delta E_\alpha$ serves as a benchmark for the non-resonant mode
energy. Equation (\ref{eq:Ealpha}) provides a simple relationship
between the actual mode energy $E_\alpha$ and the
non-resonant mode energy $\Delta E_\alpha$.

\section{Stellar Oscillation Modes and Non-Resonant Mode Energies}
\label{modes}

\begin{figure*}
\begin{center}
\includegraphics[scale=0.55]{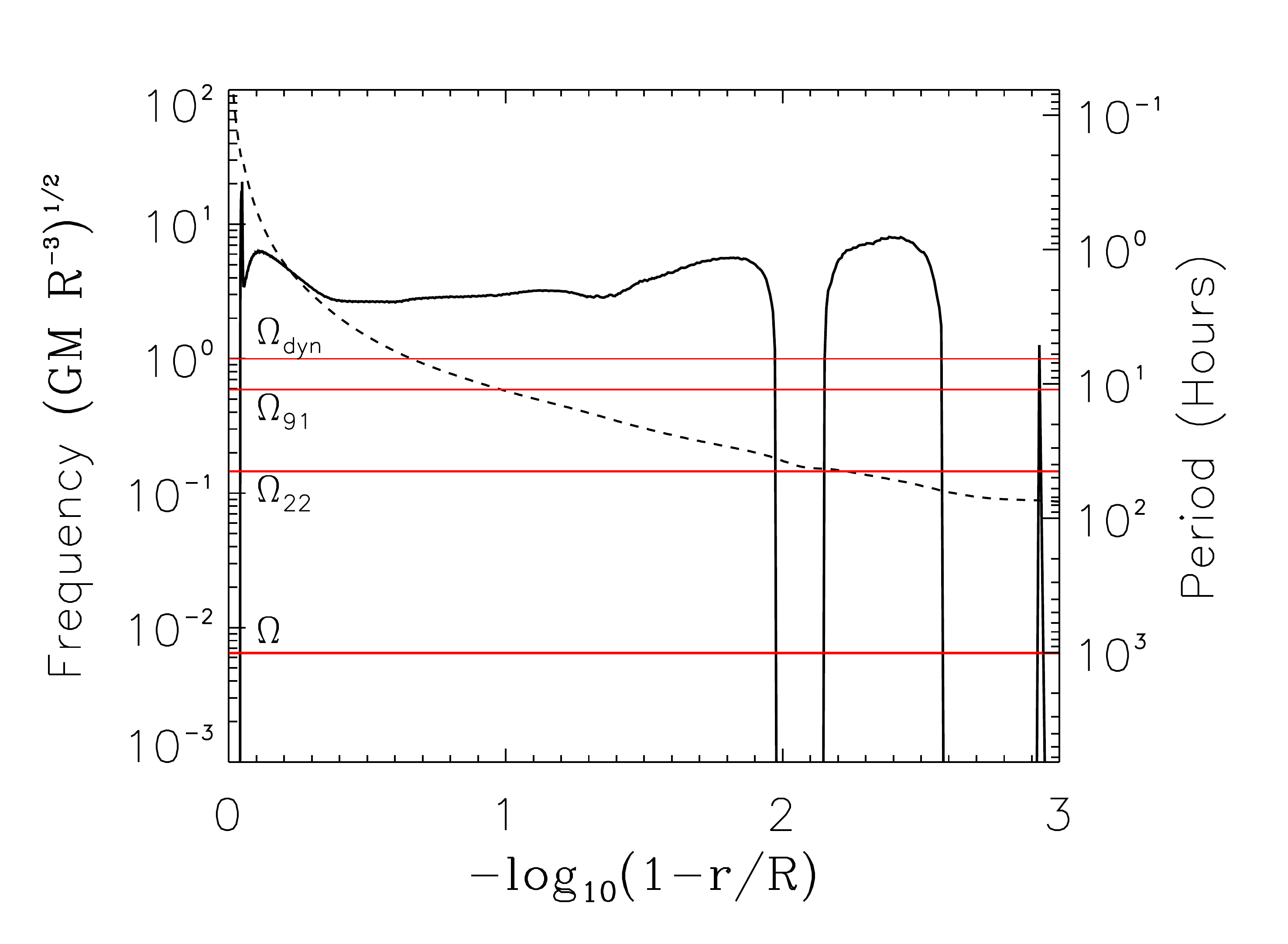}
\end{center}
\caption{ \label{plot0} Propagation diagram for our $M=2.35 M_\odot$, $R=2.34 R_\odot$ stellar model, showing the value of $N$ (solid black line) and the Lamb frequency, $L_2$ (dashed black line), in units of $(GM/R^3)^{1/2}$. The horizontal red lines denote important angular frequencies for the KOI-54 system, including (from top) the dynamical frequency of the star, $(GM/R^3)^{1/2}$; the highest frequency mode observed  in KOI-54 ($\sigma_\alpha = 91 \Omega$); the lowest frequency mode observed ($\sigma_\alpha=22.42 \Omega$); and the orbital angular frequency, $\Omega$. The $y-$axis on the right-hand side displays the corresponding periods, in units of hours.}
\end{figure*}

We construct an $M=2.35\,M_\odot$ stellar model using the 
MESA code (Paxton et al.~2010). We assume solar metallicity and evolve
the star until its radius reaches $R=2.34\,R_\odot$. These parameters are
close to star $M_2$ in KOI-54. Figure \ref{plot0} displays a propagation diagram for our stellar model. The star has a small
convective core inside radius $r=0.09R$. We make sure that the stellar model
has thermodynamically consistent pressure, density, sound speed and
Brunt-V\"as\"al\"a frequency profiles.
We have computed the $l=2$ adiabatic g-modes for this non-rotating
stellar model, including $\omega_\alpha$, $Q_\alpha$, and the mode
mass $M_\alpha\equiv
\langle\bxi_\alpha\cdot\bxi_\alpha\rangle/|\bxi_\alpha(R)|^2$. Here,
the magnitude of the surface displacement is defined by
\begin{align}
\label{bxi}
|\bxi_\alpha(R)|^2 &= \int d\Omega\, \bxi_\alpha(R) \cdot \bxi_\alpha^*(R) \nonumber \\
&= {\xi_r}_\alpha^2(R) + l(l+1){\xi_\perp}_\alpha^2(R),
\end{align}
where the $r$ and $\perp$ subscripts denote the radial and horizontal
components of the displacement vector, respectively.

We use equation (\ref{eq:dEalpha}) to compute the non-resonant mode energy 
$E_\alpha\sim\Delta E_\alpha$. The corresponding 
surface displacement $\xi_\alpha(R)$ is then obtained from
\be
|\bxi_\alpha(R)|\equiv \left({E_\alpha\over M_\alpha\sigma_\alpha^2}\right)^{1/2}.
\ee

\begin{figure*}
\begin{center}
\includegraphics[scale=0.55]{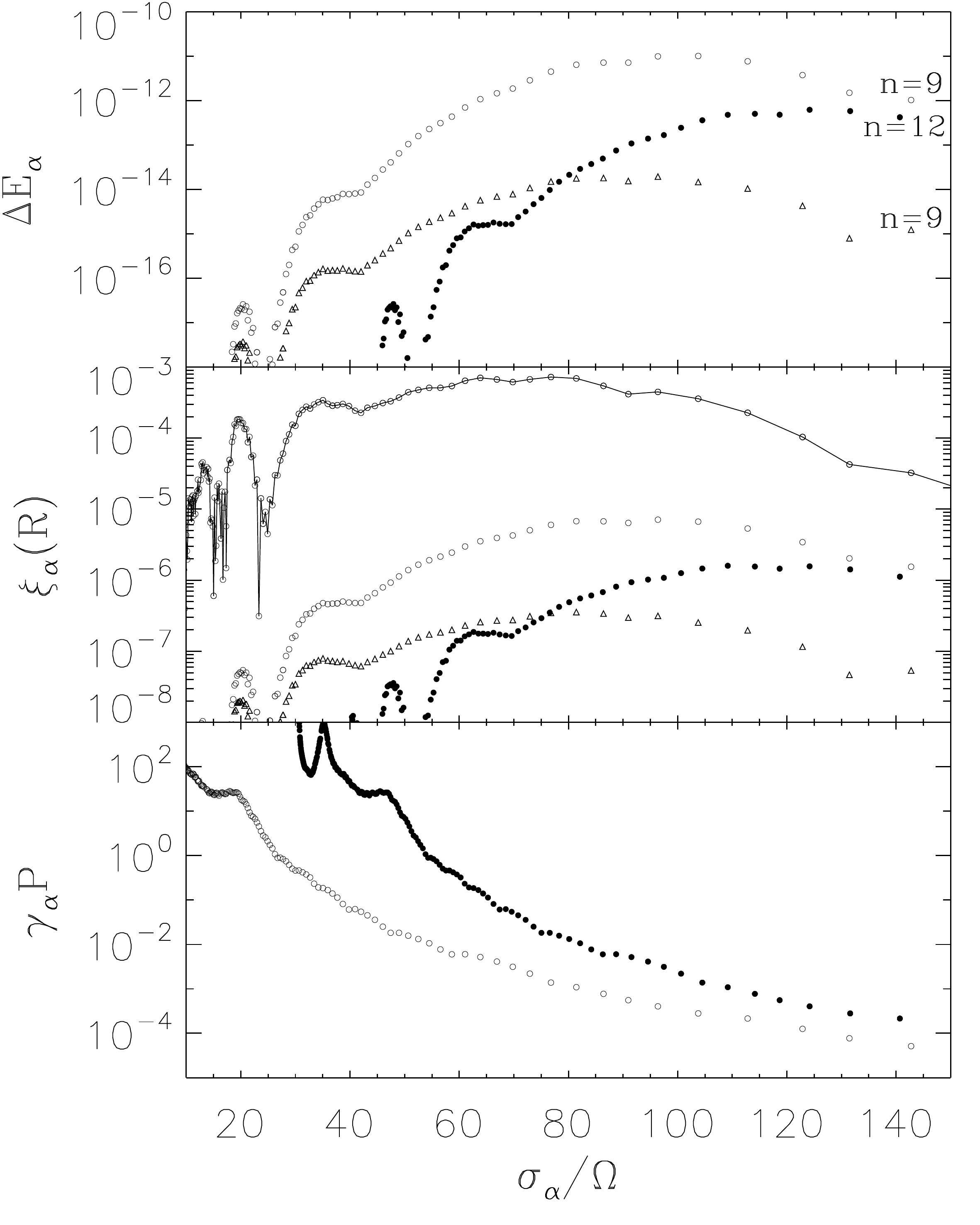}
\end{center} 
\caption{ \label{plot1} Non-resonant mode energy $\Delta E_\alpha$ (top), 
  amplitude of the radial surface displacement ${\xi_r}_\alpha(R)$
  (middle), and mode damping rates $\gamma_\alpha P$ (bottom) as a function of mode frequency $\sigma_\alpha$ (in units
  of the orbital frequency $\Omega$). The $l=m=2$ modes are plotted as
  open circles for $\Omega_s=0$ and filled circles for $\Omega_s=16.5
  \Omega$, while $m=0$ modes are plotted as triangles. The middle
  panel also displays the magnitude of the surface displacement,
  $|\bxi_\alpha(R)|$, for the $l=m=2$ modes with $\Omega_s=0$ (open
  circles connected by lines). The highest frequency modes shown are
  the $g_{9}$ mode for the $m=0$ and non-spinning $m=2$ cases, and the
  $g_{12}$ mode for the spinning $m=2$ case. The mode energy is in
  units of $GM^2/R$ and the displacement is in units of $R$.}
\end{figure*}

Figure \ref{plot1} shows the energy $\Delta E_\alpha$ and surface amplitude of
the radial component of the displacement, ${\xi_r}_\alpha(R)$, of
tidally excited modes away from resonances for the KOI-54 parameters. 
The radial displacement is directly related to the flux variation
due to the oscillation mode (see Section 4). The
most energetic modes have frequencies $80 \Omega \lo \sigma_\alpha \lo 140
\Omega$, where $\Omega$ is the orbital angular frequency, depending on
the stellar spin rate and the value of $m$. Low-order (high frequency)
modes have larger values of $Q_\alpha$ but have smaller values of
$K_{lm}(\sigma_\alpha)$, so medium-order ($n\approx 15$) modes have the largest
values of $\Delta E_\alpha$.\footnote{The orbital frequency at 
periastron is $f_p=f_\orb (1+e)^{1/2}/(1-e)^{3/2}=20.06\,f_\orb = 1/(2.084~{\rm d})$, thus $m=2$ modes with $\sigma_\alpha/\Omega\sim 40$ have the largest values of $K_{lm}$. For $m=0$, modes with $\sigma_\alpha/\Omega \sim 1$ have the largest values of $K_{lm}$.}

Figure \ref{plot1} also shows the magnitude of the displacement
$|\bxi_\alpha(R)|$ for $m=2$ modes in the zero spin limit. The total
displacement is much larger than the radial displacement for 
low-frequency modes because these modes are characterized by large
horizontal displacements and are concentrated near the surface.
Consequently, these modes have lower mode mass 
and the maximum of $|\bxi_\alpha(R)|$ shifts
to lower frequencies compared to the maximum of $\Delta E_\alpha$.

The rotation rates of the KOI-54 stars are unknown. Spectroscopic
observations constrain $V_{\rm rot}\sin i_s\lo 10$~km~s$^{-1}$,
corresponding to $V_{\rm rot}\lo 100$~km~s$^{-1}$ if the spin
inclination angle $i_s$ equals $i_s=5.5^\circ$ (Welsh et
al.~2011). This implies that the spin period $P_s\go 1.2$~days and
that $\Omega_s \lo 30 \Omega$.  Although the classical equilibrium
tide theory (e.g., Hut 1981) is not expected to be valid for our
system (Lai 1997), we can adopt the pseudosynchronous rotation
frequency $\Omega_s=\Omega_{\rm ps}=16.5\Omega$
[corresponding to $f_{ps}=1/(2.53~{\rm days})$]
as a fiducial value.

To account for the effect of stellar rotation on the tidally excited
modes, we adopt the perturbative approximation, valid 
when $\Omega_s$ is less than $\omega_\alpha^{(0)}$ (the mode frequency
in the zero-rotation limit). The mode
wave functions and $Q_\alpha$ are unchanged by the stellar rotation,
while the mode frequencies are modified according to 
$\omega_\alpha=\omega_\alpha^{(0)}-mC_{nl}\Omega_s$ and
$\sigma_\alpha=\omega_\alpha^{(0)}+m(1-C_{nl})\Omega_s$,
where $C_{nl}>0$ is a constant (e.g. Unno et al.~1989) --- our calculation 
gives $C_{nl}\simeq 0.16$ for all relevant modes.
Note that in this approximation, $\varepsilon_\alpha=\omega_\alpha^{(0)}$.
More accurate results can be obtained using the method of Lai (1997).

Stellar rotation increases the inertial-frame frequency $\sigma_\alpha$
of the $m>0$ modes, causing the dominant modes to have higher observed
frequencies. However, rotation also shifts the tidal response to
higher order g-modes with smaller values of $Q_\alpha$, so we expect
rotation to lower the mode energies.  Figure \ref{plot1} confirms that finite
(prograde) stellar rotation indeed shifts the dominant mode energy and
surface amplitude to higher-order g-modes, giving rise to smaller
$\Delta E_\alpha$ and ${\xi_r}_\alpha(R)$.

The somewhat erratic features of $\Delta E_\alpha$ and $\xi_\alpha$ as
seen in Figure \ref{plot1} for high-order (low frequency) modes are due to mode trapping effects created by the thin sub-surface convection zone in our stellar model. For these high-order modes, the tidal overlap integrals ($Q_\alpha$) depend on the precise shape of
the mode wave function, and they can be easily affected by the detailed properties
of the stellar envelope. Care must be taken in order to obtain reliable
tidal overlap integrals for these high-order modes (see Fuller \& Lai 2010).

The damping rate of a mode, $\gamma_\alpha$, can be estimated in the quasi-adiabatic limit via 
\be
\gamma_\alpha \approx \int^{r_+}_{r_-} k_r^2 \chi (\xi_r^2 + l(l+1) \xi_\perp^2) \rho r^2 dr,
\label{gamma}
\ee
where $\chi$ is the thermal diffusivity, $k_r^2 \simeq l(l+1)N^2/(r^2\omega^2)$ is the local radial wave number, and $r_+$ and $r_-$ are the boundaries of the mode's propagation cavity. Equation (\ref{gamma}) can be derived in the WKB limit from the quasi-adiabatic work function of a mode (Unno et al. 1989). \footnote{While our paper was under review, we became aware of the paper by Burkart et al.~(2011), who also used equation (\ref{gamma}) to estimate the mode damping rate. The bottom panel of Figure \ref{plot1} was added after we saw the Burkart et al. paper.} The bottom panel of Figure \ref{plot1} shows the values of $\gamma_\alpha$ calculated via equation (\ref{gamma}) for the modes of our stellar model. Lower frequency g-modes have larger damping rates due to their larger wavenumbers and because they propagate closer to the surface of the star where the diffusivity is larger. Equation (\ref{gamma}) provides an estimate for mode damping rates via radiative diffusion in the quasi-adiabatic limit; however, fully adiabatic oscillation equations must be solved for low-frequency modes for which $\gamma_\alpha$ is comparable to the mode frequency. Furthermore, modes may also damp via nonlinear processes (see Section \ref{nonlinear}).

\section{Flux Variation due to Tidally-Forced Oscillations}
\label{Resonant Modes}

In section \ref{general theory}, we considered the tidal response to
be composed of the sum of the star's natural oscillation modes,
each having its own steady-state energy $E_\alpha$.
This provides a simple relation between
the resonant mode energy and non-resonant energy.
The observed magnitude variation of KOI-54 has over 20 components with frequencies
that are almost exact multiples of the orbital frequency, i.e., they
have $\sigma_\alpha = (N+\epsilon)\Omega$ with $|\epsilon| \le 0.01$.
In this section, we examine the tidal response as a sum of oscillations
at exact multiples of the orbital frequency. Each component $U_{lm}$
of the tidal potential can be decomposed as
\be
\label{U2}
U_{lm} = - \frac{GM'W_{lm}r^l}{a^{l+1}} Y_{lm}(\theta,\phi_i) \sum_{N=-\infty}^{\infty} F_{Nm}\, e^{-iN\Omega t},
\ee
where $F_{Nm}$ is defined by the expansion 
\be
\left({a\over D}\right)^{l+1} e^{-im\Phi} = \sum_{N=-\infty}^{\infty} F_{Nm} e^{-iN\Omega t}
\ee
and is given by
\be
\label{FN}
F_{Nm} = \frac{1}{\pi} \int^\pi_0 d\Psi \frac{\cos \big[N\big(\Psi - e \sin \Psi\big) - m \Phi(t)\big]}{\big(1-e \cos \Psi\big)^{l}}.
\ee
Here, $a$ is the semi-major axis of the orbit, and $\Psi$ is the eccentric anomaly. Note that $F_{Nm}$ is
related to $K_{lm}$ [see equation (\ref{K})] by
$K_{lm}(\sigma_\alpha=N\Omega) = W_{lm} (1-e)^{l+1} \Omega^{-1}
F_{Nm}$. Inserting equation (\ref{U2}) into equation (\ref{eq:adot})
yields
\begin{align}
\label{cdot}
\dot{c}_\alpha + (i\omega_\alpha + \gamma_\alpha) c_\alpha &= \frac{i G M' W_{lm} Q_\alpha}{2 \varepsilon_\alpha a^{l+1}} \nonumber \\
&\times \sum_{N=-\infty}^{\infty} F_{Nm} e^{i(m\Omega_s -N \Omega)t},
\end{align}
whose non-homogeneous solution is
\be
\label{csol}
c_\alpha(t) = \frac{GM' W_{lm}Q_\alpha}{2\varepsilon_\alpha a^{l+1}} \sum_{N=-\infty}^{\infty} \frac{F_{Nm} e^{-i(N\Omega -m \Omega_s )t}}{(\sigma_\alpha-N\Omega) - i \gamma_\alpha}.
\ee
The total tidal response is  
$\bxi({\bf r},t) = \sum_\alpha c_\alpha(t) \bxi_\alpha({\bf r})$,
where $\bxi_\alpha({\bf r}) \propto e^{im\phi}$. When the
displacements are expressed in terms of ${\bf r}_i=(r,\theta,\phi_i)$ 
(the position vector in the inertial frame, with $\phi_i = \phi + 
m \Omega_s t$), we have
\begin{align}
\label{ctot}
\bxi({\bf r}_i,t) &= \sum_{N=-\infty}^{\infty} \sum_{\alpha} \frac{GM'W_{lm}Q_\alpha}{2 \varepsilon_\alpha a^{l+1}} \frac{F_{Nm} \bxi_\alpha({\bf r}_i)}{(\sigma_\alpha - N \Omega) - i \gamma_\alpha} e^{-iN\Omega t} \nonumber \\
&= \sum_{N=1}^{\infty} \sum_{\alpha'} \frac{GM'W_{lm}Q_\alpha}{2 \varepsilon_\alpha a^{l+1}} \bxi_\alpha({\bf r}_i) \nonumber \\
&\!\!\!\!\!\!\!\!\!\!\! \times \bigg[\frac{F_{Nm} e^{-iN\Omega t}}{(\sigma_\alpha - N\Omega) -i \gamma_\alpha} + \frac{F_{-Nm} e^{iN\Omega t}}{(\sigma_\alpha + N\Omega) -i \gamma_\alpha}\bigg] + c.c.
\end{align}
where $c.c.$ denotes the complex conjugate, and $\sum_{\alpha'}$
implies that the sum is restricted to modes with $\omega_\alpha >0$
(including both $m>0$ and $m<0$ modes) (by contrast, $\sum_\alpha$
includes both $\omega_\alpha >0$ and $\omega_\alpha<0$ terms, as well
as both $m>0$ and $m<0$).  We have omitted the $N=0$ term for
simplicity because it is not part of the dynamical response. Each
$\bxi_N$ (oscillating at frequency $N\Omega$) is then a sum over the
star's oscillation modes, with large contributions
coming from nearly-resonant modes with $\sigma_\alpha \approx N \Omega$.

We use the result of Buta \& Smith (1979) (see also Robinson
et al.~1982) to estimate the magnitude variation $(\Delta \rm{mag})$
associated with each component of the tidal response. The magnitude
variation of the star has two primary components: a geometrical
component, $(\Delta {\rm mag})_G$, due to distortions in the shape of
the star, and a temperature component, $(\Delta {\rm mag})_T$, due to
perturbations in the surface temperature of the star's photosphere.

For an $l=m=2$ mode with surface radial displacement $\xi_{r\alpha}(R)$,
the amplitudes of the bolometric magnitude variations are
\begin{align}
\label{deltam2T}
(\Delta {\rm mag})_{\alpha T}^{(m=2)} & \simeq -1.7\,\gamma_l \sin^2 i_s\, \frac{\Gamma_2-1}{\Gamma_2} \nonumber \\
& \times \bigg[\frac{l(l+1)}{\hat{\omega}_\alpha^2} - 4 - \hat{\omega}_\alpha^2\bigg] \frac{{\xi_r}_\alpha(R)}{R} 
\end{align}
and
\be
\label{deltam2G}
(\Delta {\rm mag})_{\alpha G}^{(m=2)} \simeq -0.42\,(\alpha_l + \beta_l)
\sin^2 i_s\, \frac{{\xi_r}_\alpha(R)}{R}.
\ee
Here, $\Gamma_2 \simeq 5/3$ is the adiabatic index of gas at the
surface of the star, $\hat\omega=\omega_\alpha/\sqrt{GM/R^3}$ is the
dimensionless mode frequency, and $\gamma_l \approx 0.3$ and
$\alpha_l+\beta_l \approx -1.2$ are bolometric limb darkening
coefficients appropriate for A stars. 
For KOI-54, the value of the spin inclination angle $i_s$ 
(the angle between the spin axis and the line of sight) is unknown,
but we may use the system's orbital 
inclination of $i_{\rm orb} = 5.5^\circ$ as a first guess,
although it is possible that the star's spin axis is inclined
relative to the orbital angular momentum axis.
Similarly, for $l=2$, $m=0$ modes, the amplitudes of the magnitude 
variations are
\begin{align}
\label{deltam0T}
(\Delta {\rm mag})_{\alpha T}^{(m=0)} &\simeq -1.4 \gamma_l\,\big(3 \cos^2 i_s -1\big)\,\frac{\Gamma_2-1}{\Gamma_2} \nonumber \\
&\times \bigg[\frac{l(l+1)}{\hat{\omega}_\alpha^2} - 4 - \hat{\omega}_\alpha^2 \bigg] \frac{{\xi_r}_\alpha(R)}{R}
\end{align}
and
\be
\label{deltam0G}
(\Delta {\rm mag})_{\alpha G}^{(m=0)} \simeq -0.34\,
(\alpha_l + \beta_l) \big(3 \cos^2 i_s -1\big) \frac{{\xi_r}_\alpha(R)}{R}.
\ee

In equations (\ref{deltam2T}) and (\ref{deltam0T}), the factor of
$[l(l+1)/\hat\omega_\alpha^2 - 4 -\hat\omega_\alpha^2]$ arises from 
the outer boundary condition formulated by Baker \& Kippenhahn (1965)
(see also Dziembowski 1971),
in which the radial derivative of the Lagrangian pressure perturbation vanishes
at the outer boundary (in our case, the stellar photosphere), such that
the mode energy vanishes at infinity in an isothermal atmosphere. It
is not clear how well this boundary condition applies for a real star,
nor how it should be modified when non-adiabatic effects are taken into 
account. For the
$g$-modes considered here, the factor $l(l+1)/\hat\omega_\alpha^2 \gg 1$,
causing the temperature effect to overwhelm the geometrical effect for
magnitude variations.

The magnitude variation for each frequency $N\Omega$ is a sum of the 
variations due to individual modes:
\begin{align}
\label{deltamN}
(\Delta {\rm mag})_N = \sum_{\alpha'} \frac{GM'W_{lm}Q_\alpha a_\alpha}{2 \varepsilon_\alpha a^{l+1}} \frac{{\xi_r}_\alpha(R)}{R} \nonumber \\
\times \bigg[\frac{F_{Nm} e^{-iN\Omega t}}{(\sigma_\alpha - N\Omega) -i \gamma_\alpha} + \frac{F_{-Nm} e^{iN\Omega t}}{(\sigma_\alpha + N\Omega) -i \gamma_\alpha}\bigg] + c.c.,
\end{align}
where $a_\alpha$ is the constant in front of ${\xi_r}_\alpha(R)/R$ in
equations (\ref{deltam2T})-(\ref{deltam0G}), and $N>0$.

\begin{figure*}
\begin{center}
\includegraphics[scale=0.65]{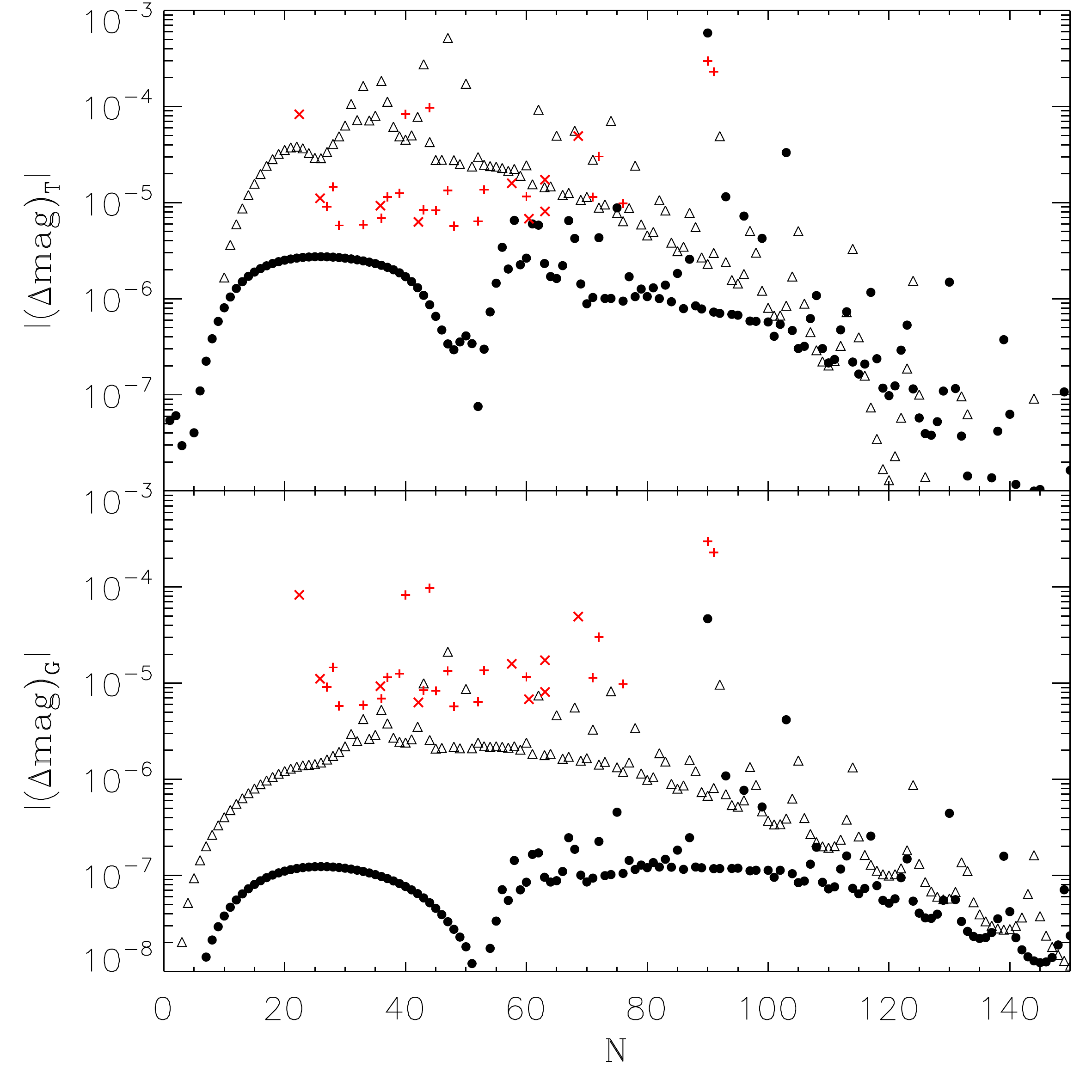}
\end{center}
\caption{ \label{plot2} Magnitude variation $\Delta {\rm mag}_N$ as a function of $N$
  due to the temperature effect (top) and geometrical effect (bottom)
  for $l=m=2$ modes (filled circles) and $l=2$, $m=0$ modes
  (triangles). We also plot the observed magnitude variations that are
  an integer multiple of the orbital frequency (red plus symbols), and
  those that are not an integer multiple of the orbital frequency (red
  x symbols). The plot uses $i_s=10^{\circ}$ and $\Omega_s \simeq 15
  \Omega$. }
\end{figure*}

Using the values of $\varepsilon_\alpha$, $Q_\alpha$,
$\bxi_\alpha(r)$, and $\gamma_\alpha$ calculated in Section \ref{general theory}, we compute
each term $(\Delta {\rm mag})_N$ in equation (\ref{deltamN}). Figure
\ref{plot2} shows a plot of the magnitude variation as a function of
$N$, along with the observed magnitude variations in KOI-54 as
determined by Welsh et al. (2011). To make this plot, we have
subtracted out the contribution from the 
\textquotedblleft equilibrium
tide'', because the equilibrium tide is responsible for the periodic
brightening of KOI-54 near periastron and was subtracted out by Welsh
et al. (2011) in order to obtain the magnitude variations due to
tidally-induced stellar oscillations. 
The equilibrium tide can be computed by taking the
limit $\sigma_\alpha \gg N \Omega$, i.e., by setting $N \Omega=0$
for each term inside the sum in equation (\ref{deltamN}). 
We adopt a spin inclination angle $i_s=10^\circ$. 

Figure \ref{plot2} includes the magnitude variations due to the
temperature effect and those due to geometrical effects.
For the adopted spin inclination ($10^\circ$), the $m=0$
modes dominate the observed magnitude variations, although a
nearly-resonant $m=2$ mode can be visible. However, it is 
obvious from Figure \ref{plot2} that our computed
variations due to the temperature effect of
$m=0$ modes are appreciably larger than those observed. 
In fact, the magnitude variations due to
geometrical effects reproduce the observed variations much more
accurately. The over-prediction of the magnitude variations 
from the temperature effect is mostly likely due to non-adiabatic effects in the stellar atmosphere, which
renders equations (\ref{deltam2T}) and (\ref{deltam0T}) inaccurate
for low-frequency modes. Indeed, Buta \& Smith (1979) also found that for main sequence B stars,
the predicted magnitude variations due
to the temperature effect for low-frequency modes were much larger than
what was observed, and they speculated that the mismatch was due to
non-adiabatic effects in the outer layers of the star. 

Our results depicted in Figure \ref{plot2}
suggest that most of the observed magnitude
variations (with the exception the highest-frequency modes)
in KOI-54 are due the geometrical surface distortions produced by $m=0$ 
modes that happen to be nearly resonant with a harmonic of the orbital
frequency. The actual magnitude variations due to the temperature 
effect (when non-adiabatic effects are properly taken into account
near the stellar photosphere) should not be much larger than the geometric effect.
Note that in Figure \ref{plot2} we have plotted the absolute values of the
magnitude variations for the temperature and geometrical effects, but
that the temperature and geometrical effects have opposite sign and
tend to cancel each other out.

To produce the highest-frequency modes ($90f_{\rm orb}$ and 
$91f_{\rm orb}$) observed in KIO-54, in Figure \ref{plot2} we have fine-tuned
the spin of the star to $\Omega_s\simeq 15\Omega$ such that the $N=90$
oscillation is nearly resonant with one of the star's oscillation
modes. In section 5
we argue that the $N=90$ and $N=91$ oscillations are likely due to an
$m=2$ mode in each star that is locked in resonance. The fine-tuning
produces a magnitude variation due to the temperature effect that is
comparable to one of the observed oscillations, although the
predicted geometrical magnitude variation is somewhat less than what
is observed. If the $N=90$ and $N=91$ oscillations are indeed due to
resonant $m=2$ modes, then it is likely that the predicted magnitude
variation due to the temperature effect (based on adiabatic
approximation) is more accurate for these modes due to their higher
frequencies. This is not unreasonable since non-adiabatic effects in
the stellar photosphere are expected to be less important for
low-order (high frequency) modes, a point also emphasized by Buta \&
Smith (1979). In any case, a more in-depth analysis of the non-adiabatic
oscillation modes (particularly their flux variations)
for the stars of the KOI-54 system is needed.

\section{Secular Spin-Orbit Evolution and Resonance Locking}
\label{resonancelocking}

In the previous sections we proposed that a resonance with 
$\sigma_\alpha\sim 90\Omega$ creates the largest observed magnitude variations in KOI-54. Here we study the secular evolution of the stellar 
spin and binary orbit, and how resonances may naturally arise during the evolution. Several aspects of tidal resonance locking in massive-star binaries were previously explored by Witte \& Savonije (1999, 2001).

Dynamical tides lead to spin and orbital evolution, with the orbital
energy and angular momentum evolving according to
\be
\label{dEdJ}
\dot E_\orb=-\sum_\alpha 2\gamma_\alpha E_\alpha,
\quad
\dot J_\orb=-\sum_\alpha 2\gamma_\alpha E_\alpha \frac{m}{\sigma_\alpha},
\ee
where we have used the fact that the mode angular momentum (in the inertial
frame) is related to the mode energy by $J_\alpha=(m/\sigma_\alpha)E_\alpha$.
The orbital frequency $\Omega$ therefore evolves as 
\be
\frac{\dot\Omega}{\Omega}=\sum_\alpha \frac{f_\alpha}{ t_{d\alpha}},
\ee
where 
\be
f_\alpha=\left[4\sin^2(\pi\sigma_\alpha/\Omega)+(\gamma_\alpha P)^2
\right]^{-1},
\ee
and $t_{d\alpha}^{-1}$ specifies
the orbital decay rate due to a single non-resonant mode $\alpha$:
\begin{align}
t_{d\alpha}^{-1}&=\frac{3\gamma_\alpha\Delta E_\alpha}{|E_{\rm orb}|}\nonumber \\
&=\frac{12\pi^2\gamma_\alpha}{(1-e)^6}
\left(\frac{M'}{ M}\right)\left(\frac{R}{ a}\right)^5
\left(\frac{\sigma_\alpha}{\varepsilon_\alpha}\right)|Q_\alpha K_{lm}|^2.
\end{align}
Using the KOI-54 parameters and assuming $\sigma_\alpha\sim\varepsilon_\alpha$,
we have
\be
t_{d\alpha}^{-1} \simeq 8.9\times 10^{-10}\gamma_\alpha \left(\frac{Q_\alpha K_{lm}}{10^{-4}}\right)^2.
\ee
Resonance occurs when a mode has frequency $\sigma_\alpha=N\Omega$ for an integer value of $N$.

\subsection{Need for Resonance Locking}

We first consider how likely it is to observe a mode near resonance when
$\sigma_\alpha$ is held constant
(i.e., it does not change during the period of
resonance-crossing). 

As can be seen from Figure \ref{plot1}, the $m=2$ modes dominate tidal energy transfer in the KOI-54 system, yet (as evidenced from Figure \ref{plot2}), most of the visible modes (except $N=90,91$) are $m=0$ modes. Consider a particular $m=0$ mode near resonance, with $\sigma_\alpha=(N+\epsilon)\Omega$ and $|\epsilon|\ll 1$. Since the mode contributes little to the tidal energy transfer, the probability of being close to resonance ($|\epsilon|<\epsilon_0$) is simply $P_{\rm res} \simeq 2 \epsilon_0$. Figure \ref{plot2} indicates that an $m=0$ mode with $25 \lo N \lo 80$ requires $|\epsilon_0| \lo 0.1$ to be visible [mode visibility scales as $|(\Delta {\rm mag})| \propto 1/|\epsilon|$]. In each star there are about forty $m=0$ modes in this frequency range, thus we should expect about 8 of these modes from each star to be visible, in rough agreement with Figure \ref{plot2} and the observations. 

Now let us consider modes that significantly influence the tidal energy transfer (these include $m=2$ modes such as the $N=90,91$ modes, but may also include $m=0$ modes very near resonance -- if they occur). For the KOI-54 system, the modes that contribute significantly to the tidal energy transfer have $\sigma_\alpha\go 2 \Omega_p\simeq 40\Omega$, thus $N=\sigma_\alpha/\Omega \go 40$. 
Consider a particular mode near resonance, and suppose that the tidal energy transfer is dominated by the 
resonant mode ($\alpha$). The orbital decay rate during the resonance
is given by $\dot P/P=-t_{d\alpha}^{-1}
\left[(2\pi\epsilon)^2+(\gamma_\alpha P)^2\right]^{-1}$.
Thus the time that the system spends ``in resonance'' 
($|\epsilon|<\epsilon_0$) is $(\Delta t)_{\epsilon_0}=t_{d\alpha} 
(2\epsilon_0/N)\left[(2\pi\epsilon_0)^2/3+
(\gamma_\alpha P)^2\right] \sim 8 \pi^2 t_{d \alpha} |\epsilon_0|^3 /(3N)$. By contrast, the time it takes the orbit
to evolve between resonances (for the same mode) is $(\Delta t)_{\rm nr}
\sim t_{d}/N$, where $t_d$ is the orbital decay timescale due to all
non-resonant modes ($t_d$ may be a factor of $\sim 10$ smaller than $t_{d\alpha}$). The probability of seeing a mode very near resonance is thus
\be
P_{\rm res} = \frac{(\Delta t)_{\epsilon_0}}{(\Delta t)_{\rm nr}} \simeq \frac{8 \pi^2}{3} \frac{t_{d \alpha}}{t_d} |\epsilon_0|^3.
\label{probability}
\ee
Figure \ref{plot2} indicates that (for in inclination of $i_s=10^\circ$) we require $|\epsilon_0| \lo 10^{-2}$ for an $m=2$ mode to be visible, for which the probability is $P_{\rm res} \lo 3\times10^{-4}$. Thus, at first glance, the chances of observing even a single tidally-excited $m=2$ mode in the KOI-54 system are slim. The $N=90,91$ modes require $|\epsilon_0| \lo 10^{-2}$, even if they are produced by $m=0$ modes. It is therefore extremely unlikely to observe the system with such large amplitude modes, unless (by some mechanism) they are locked into resonance. In the sections that follow, we outline such a resonance locking mechanism and how it applies to the KOI-54 system.

\subsection{Critical Resonance-Locking Mode}

We now consider the possibility of resonance locking for the $m=2$ modes.
Tidal angular momentum transfer to the star 
increases the stellar spin $\Omega_s$, thereby changing 
the mode frequency $\sigma_\alpha$. There exists a particular 
resonance, $\sigma_\alpha=N_c\Omega$, for which $\sigma_\alpha$ and
$\Omega$ change at the same rate, i.e.,
\be
{\dot\sigma_\alpha\over \sigma_\alpha}
={\dot\Omega\over \Omega},
\ee
so that the resonance can be maintained once it is reached.
With $\dot J_s=-\dot J_\orb$ and the mean stellar rotation
rate $\bar\Omega_s=J_s/I$ (where $I$ is the moment of inertia), we have
$\dot\sigma_\alpha=mB_\alpha\dot J_s/I$,
where $B_\alpha\equiv (d\sigma_\alpha/d\bar\Omega_s)/m$.
Assuming that a single resonant mode dominates the tidal energy and angular
momentum transfer, we find $\dot\sigma_\alpha/\sigma_\alpha
=-m^2B_\alpha\dot E_\orb/(IN^2\Omega^2)$. Thus
\be
\left({\dot\sigma_\alpha\over \sigma_\alpha}\right)_{\rm tide}
=\left(\frac{N_c}{ N}\right)^2\left({\dot\Omega\over\Omega}\right)_{\rm tide},
\label{eq:rates}\ee
where 
\be
\label{nc}
N_c=m\left(\frac{B_\alpha\mu a^2}{ 3I}\right)^{1/2}
\ee
(with $\mu=MM'/M_t$ the reduced mass of the binary) 
corresponds to the critical resonance-locking mode.
The subscript ``tide'' in equation (\ref{eq:rates}) serves as a reminder
that the changes of $\sigma_\alpha$ and $\Omega$ are due to the tidal
interaction.

For the KOI-54 parameters (with $a\simeq 39R$ and $m=2$), we have
\be
\label{nc2}
N_c \simeq 131\left({B_\alpha\over 0.84}\right)^{1/2}
\left({\kappa\over 0.05}\right)^{-1/2},
\ee
where $\kappa=I/(MR^2)$. If we assume that the star maintains rigid-body 
rotation during tidal spin up, then $B_\alpha=1-C_{nl}$. Numerical
calculation for our stellar model gives $B_\alpha\simeq 0.84$ 
(with very weak dependence on modes) and $\kappa\simeq 0.040$, giving $N_c \simeq 146$; 
a less evolved star would have $\kappa \simeq 0.047$, giving $N_c \simeq 135$. Note that in the above we
consider only tides in one star ($M$) -- when tides in the other star
are considered, $\dot\Omega$ would be larger and the effective value of $N_c$ would be reduced.
If identical resonances occur in the two stars, the effective value of $N_c$ would be
reduced by a factor of $\sqrt{2}$ to $N_c \simeq 92$ (for $\kappa =0.05$). See section 5.3.3 and equations (\ref{nc_1})-(\ref{nc_2}) below.

The above consideration assumes that the spin evolution of the star is
entirely driven by tide. In reality, the star can experience intrinsic
spindown, either due to a magnetic wind or due to radius expansion
associated with stellar evolution. We may write
$\dot\Omega_s=(\dot\Omega_s)_{\rm tide}-|\dot\Omega_s|_{\rm sd}$,
where the second term denotes the contribution due to the intrinsic
spindown. The corresponding rate of change for $\sigma_\alpha$ is then
\be
{\dot\sigma_\alpha\over\sigma_\alpha}=
\left({\dot\sigma_\alpha\over\sigma_\alpha}\right)_{\rm tide}
-\left|{\dot\sigma_\alpha\over\sigma_\alpha}\right|_{\rm sd}.
\label{eq:sigdot}\ee
On the other hand, the orbital evolution remains solely driven by tides:
\be
{\dot\Omega\over\Omega}=\left({\dot\Omega\over\Omega}\right)_{\rm tide}
=\left(\frac{N}{N_c}\right)^2
\left({\dot\sigma_\alpha\over\sigma_\alpha}\right)_{\rm tide}.
\label{eq:Omegadot}\ee
Comparing equations (\ref{eq:sigdot}) and (\ref{eq:Omegadot}), we see 
that {\it $N_c$ represents the upper boundary for resonance locking:}
for $N>N_c$, resonance locking ($\dot\sigma_\alpha/\sigma_\alpha
=\dot\Omega/\Omega$) is not possible. For $N<N_c$, resonance locking can be
achieved: the value of $(\dot\sigma_\alpha/\sigma_\alpha)_{\rm tide}$
depends on the closeness to the resonance, which in turn depends on 
the intrinsic spindown timescale of the star [see equations 
(\ref{eq:falpha}) and (\ref{eq:epsiloneq}) below].

The fact that $N_c$ naturally falls in the range close to $N=90$ and
$N=91$, the most prominent modes observed in KOI-54, is encouraging. 
In the next subsection, we consider how the system may naturally evolve into
a resonance-locking state with $N<N_c$.

\subsection{Evolution Toward Resonance}

As noted above, even in the absence of tides, the star can experience spin-down,
either due to magnetic wind or due to radius expansion associated
with stellar evolution. Furthermore, the intrinsic mode frequencies will change as the internal structure of the star changes due to stellar evolution. We now show that these effects can naturally lead to evolution of the system toward resonance locking.

The evolution equation for the stellar spin reads
$\dot\Omega_s=-(\dot J_{\rm orb}/I)-(\Omega_s/t_{\rm sd})$,
where $t_{\rm sd}$ is the ``intrinsic'' stellar spin-down time scale
associated with radius expansion and/or magnetic braking. Using equation (\ref{dEdJ}), 
we have
\be
\label{doms}
{\dot\Omega_s\over\Omega_s}=\sum_\alpha\left({m\mu a^2\over 3I}
\right)\left({\Omega^2\over\Omega_s\sigma_\alpha}\right)
{f_\alpha\over t_{d\alpha}}-{1\over t_{\rm sd}}.
\ee
From equation (\ref{dEdJ}), we find that the orbital frequency and eccentricity 
evolve according to
\be
\label{dom}
{\dot\Omega\over\Omega}=\sum_\alpha{f_\alpha\over t_{d\alpha}},
\ee
\be
\label{de}
{e\,{\dot e}\over 1-e^2}=-{1\over 3}\sum_\alpha {f_\alpha\over t_{d\alpha}} \left[1-{m\Omega\over \sigma_\alpha (1-e^2)^{1/2}}\right].
\ee
To leading order in $\Omega_s$, the mode frequency depends on $\Omega_s$ 
via 
\be
\label{dsig}
\sigma_\alpha=\omega_\alpha^{(0)}+mB_\alpha\Omega_s.
\ee
For simplicity, we assume
that the secular change of $\sigma_\alpha$ is only caused
by the $\Omega_s$-evolution (e.g., we neglect the variation of
$\omega_\alpha^{(0)}$ due to stellar evolution -- this effect can be absorbed
into the spindown effect on the mode; see Witte \& Savonije 1999). 

\subsubsection{Single Mode Analysis}

To gain some insight into the evolutionary behavior of the system,
we first consider the case where one of the modes is very
close to resonance, i.e., 
\be
\sigma_\alpha=(N+\epsilon)\Omega,
\label{epsilon}
\ee
while all the other modes are non-resonant. We then write the
orbital and spin frequency evolution equations as
\be
\label{dom2}
{\dot\Omega\over\Omega}={f_\alpha\over t_{d\alpha}}+{1\over t_d},
\ee
\be
\label{doms2}
{\dot\Omega_s\over\Omega_s}={N_c^2\over mB_\alpha}{\Omega^2\over
\Omega_s\sigma_\alpha}{f_\alpha\over t_{d\alpha}}-{1\over t_{sd}}
+{1\over t_{su}}.
\ee
Here $t_d^{-1}$ and $t_{su}^{-1}$ are the orbital decay rate and
spin up rate due to all the non-resonant modes.\footnote{For simplicity,
here we do not consider the possibility where the star rotates so fast that 
dynamical tides spindown the star -- such possibility occurs when the
star rotates at a rate somewhat beyond $\Omega_p$ (see Lai 1997).}
They are approximately 
given by
\begin{align}
& t_{d}^{-1}\simeq \sum_\alpha 3\gamma_\alpha {\Delta E_\alpha\over 
|E_{\rm orb}|}\sim {3 \gamma\Delta E\over |E_{\rm orb}|},\\
& t_{su}^{-1}\simeq {1\over I\Omega_s}\sum_\alpha 2\gamma_\alpha 
{m\over\sigma_\alpha}\Delta E_\alpha\sim {2\gamma \Delta J\over I\Omega_s},
\end{align}
where $\Delta E=\sum_\alpha\Delta E_\alpha$ and $\Delta J=\sum_\alpha
(m/\sigma_\alpha)\Delta E_\alpha$ are the energy and angular momentum 
transferred from the orbit to the star in the ``first'' periastron passage
(see Section \ref{general theory}), and $\gamma$ is the characteristic mode damping rate
of the most important modes in the energy transfer. Since 
$\Delta J\sim \Delta E/\Omega_p$ (where $\Omega_p$ is the orbital
frequency at periastron), we find
\be
{t_{su}^{-1}\over t_d^{-1}}\sim {\mu a^2\over 3I}{\Omega^2\over
\Omega_s\Omega_p}.
\ee
Note that $t_d^{-1}$ can be a factor of a few ($\sim 10$) larger than $t_{d\alpha}^{-1}$, and $t_{su}^{-1}$ is a factor of a few larger than $t_d^{-1}$.

Equations (\ref{epsilon})-(\ref{doms2}) can be combined to yield the evolution equation for
$\epsilon=\sigma_\alpha/\Omega-N$:
\be
{\dot\epsilon\over N}=\left[\left({N_c\over N}\right)^2-1\right]
{f_\alpha\over t_{d\alpha}}-{1\over t_D},
\label{eq:epdot}\ee
where 
\be
{1\over t_D}={mB_\alpha\Omega_s\over N\Omega}\left({1\over t_{sd}}
-{1\over t_{su}}\right)+{1\over t_d}.
\ee
Equation (\ref{eq:epdot}) provides the key for understanding the condition
of achieving mode locking:

(i) For $N>N_c$: The RHS of Equation (\ref{eq:epdot}) is always negative, and
the system will pass through the resonance ($\epsilon=0$) without locking.
Physically, the reason is that for $N>N_c$, the orbit decays
faster (during resonance) than the mode frequency can catch up,
so the system sweeps through the resonance.

(ii) For $N<N_c$ and $t_D^{-1}\gg t_{d\alpha}^{-1}$: Starting from a
non-resonance initial condition ($\epsilon_{\rm in}\sim 0.5$, or 
$f_\alpha\sim 1$), the system will evolve toward a resonance-locking state
($\dot\epsilon=0$), at which
\be
f_\alpha={t_{d\alpha}\over \delta_N t_D}\gg 1
\label{eq:falpha}
\ee
where $\delta_N\equiv (N_c/N)^2-1$. The ``equilibrium'' value of
$\epsilon$ is given by 
\be
\epsilon_{\rm eq} \simeq {1\over 2\pi}\left[{t_D\delta_N\over t_{d\alpha}}
-(\gamma_\alpha P)^2\right]^{1/2}\ll 1.
\label{eq:epsiloneq}
\ee
This resonance-locking state can be achieved when
\be
\label{saturation}
t_D\delta_N/t_{d\alpha}>(\gamma_\alpha P)^2,
\ee
as otherwise resonance ``saturation'' [$f_\alpha\le 
(\gamma_\alpha  P)^{-2}$] occurs, and the system will sweep through 
the resonance. Note that during the resonance-locking phase, 
the stellar spin increases as
\be
\label{domsres}
{\dot\Omega_s\over\Omega_s}={1\over\delta_N}
\left(t_{sd}^{-1}-t_{su}^{-1}\right)+{N_c^2\Omega\over
N\delta_N mB_\alpha\Omega_s} t_d^{-1},
\ee
and the orbital frequency increases as
\be
\label{domres}
\frac{\dot{\Omega}}{\Omega} = \frac{mB_\alpha}{N\delta_N}\frac{\Omega_s}{\Omega}\Big(t_{sd}^{-1} - t_{su}^{-1}\Big) + \frac{N_c^2}{N^2\delta_N}t_d^{-1}.
\ee

The above analysis assumes $t_D^{-1}>0$. In the absence of the
intrinsic stellar spin-down (i.e., $t_{sd}^{-1}=0$), we have
\be
t_D^{-1}\sim -{N_c^2\over mN}\left({20\Omega\over\Omega_p}\right)
t_d^{-1}+t_d^{-1}.
\ee
Thus $t_D$ can be negative for $N\lo N_c^2/40$. In this case,
one may expect mode-locking for $N>N_c$. Nevertheless, $t_D^{-1}$ is 
only moderately larger (a factor of 10 or so) than $t_{d\alpha}^{-1}$
without intrinsic spin-down, so a close resonance with $f_\alpha\gg1$
(or $|\epsilon|\ll 1$) is unlikely.

\subsubsection{Resonance Locking for $m=0$ Modes}
\label{m0}

Above (and in other sections of this paper), we considered the frequency of a mode in the inertial frame to first order in $\Omega_s$: $\sigma_\alpha = \varepsilon_\alpha + m B_\alpha \Omega_s$. In this approximation, $m=0$ modes have constant frequency and thus cannot lock into resonance. However, Lai (1997) find that to second order in $\Omega_s$, the frequency of $m=0$ modes can be approximated by 
\be
\label{sigm0}
\frac{\sigma_{\alpha,m=0}}{\varepsilon_\alpha} \simeq 1 + \frac{6}{7}\bigg(\frac{\Omega_s}{\varepsilon_\alpha}\bigg)^2.
\ee
Recall that $\varepsilon_\alpha$ is the mode frequency in the absence of rotation. Thus, to second order, the frequency of $m=0$ modes can change and resonance locking is possible.

Let us examine the scenario in which an $m_\alpha=2$ mode with $N_{m=2}\simeq90$ is locked in resonance, as we suspect is the case for KOI-54. Assuming this mode dominates the tidal interaction, the system evolves such that 
\begin{align}
\label{dm2}
N_{m=2} \dot{\Omega} &= \dot{\sigma}_{\alpha,m=2} \nonumber \\
&= m_\alpha B_{\alpha} \dot{\Omega}_s,
\end{align}
Then there will be a value of $N_{m=0}$ for an $m=0$ mode such that 
\be
\label{dm0}
N_{m=0} \dot{\Omega} = \dot{\sigma}_{\alpha,m=0}.
\ee
Using equation (\ref{dm2}) for the LHS of equation (\ref{dm0}), equation (\ref{sigm0}) for the RHS of equation (\ref{dm0}), and the condition $\sigma_{\alpha,m=0} = N_{m=0} \Omega$, we find 
\be
\label{Nm0}
N_{m=0} = \bigg[\frac{12 N_{m=2}}{7 m_\alpha B_\alpha \big[\Omega/\Omega_s - m_\alpha B_\alpha/(2 N_{m=2})\big]}\bigg]^{1/2}.
\ee
Using $N_{m=2}=90$, $m_\alpha=2$, $B_\alpha=0.84$, and $\Omega_s=\Omega_{ps}= 16.5 \Omega$, we find that $N_{m=0} \simeq 42$. Thus, an $m=0$ mode with $\sigma_\alpha \simeq 42 \Omega$ may be able to stay close to resonance for an extended period of time. KOI-54 has two very visible modes at $\sigma_3=44 \Omega$ and $\sigma_4=40 \Omega$. We speculate that these modes may correspond to an $m=0$ mode in each star that is nearly locked in resonance due to the orbital evolution produced by a locked $m=2$ mode with $N_{m=2}\simeq 90$. However, because the value of $N_{m=0}$ depends on $\Omega_s$ (which is currently unknown for KOI-54), we will not consider $m=0$ mode locking in the remainder of this paper.

\subsubsection{Resonance Locking in Both Stars}

In Section 5.3.1, we considered resonant locking of an $m=2$ mode
in the primary star $M$. Since the two stars in the KOI-54 system are
quite similar, resonant locking may be achieved in both stars simultaneously.
We now consider the situation in which an $m=2$ 
mode $\sigma_\alpha$ in star $M$ and an $m'=2$ 
mode $\sigma_{\alpha'}$ in star $M'$ are both very 
close to orbital resonance, i.e.,
\be
\sigma_\alpha=(N+\epsilon)\Omega,\qquad
\sigma_{\alpha'}=(N'+\epsilon')\Omega,
\ee
while all other modes (in both stars) are non-resonant. Here, all primed quantities refer to star $M'$, and unprimed 
quantities refer to star $M$.
The orbital evolution equation then reads
\be
{\dot\Omega\over\Omega}={f_\alpha\over t_{d\alpha}}
+{f_{\alpha'}\over t_{d\alpha'}}+{1\over t_d},
\ee
while the spin evolution is governed by equation (\ref{domsres}) for star $M$ and a
similar equation for $M'$. We then find
\begin{align}
&{\dot\epsilon\over N}=\left[\left({N_c\over N}\right)^2-1-x\right]
{f_\alpha\over t_{d\alpha}}-{1\over t_D},\\
&{\dot\epsilon'\over N'}=\left[\left({N_c'\over N'}\right)^2-1-{1\over x}\right]
{f_{\alpha'}\over t_{d\alpha'}}-{1\over t_D'},
\end{align}
where 
\be
x\equiv {f_{\alpha'}/t_{d\alpha'}\over
f_\alpha/t_{d\alpha}}={\dot E_{\alpha'}\over \dot E_\alpha},
\ee
with $\dot E_\alpha$ and $\dot E_{\alpha'}$ the energy dissipation rates
(including resonances) due to the resonant modes in star $M$ and $M'$, 
respectively.
Thus, analogous to Section 5.3.1, for $t_D>0$ and $t_D'>0$,
the necessary conditions for resonant 
mode locking in both stars are 
\be 
N<N_{c,{\rm eff}},\quad {\rm and}\quad N'<N_{c,{\rm eff}}',
\ee
with
\be
\label{nc_1}
N_{c,{\rm eff}}={N_c\over\sqrt{1+x}}=m\left[{B_\alpha\mu a^2\over 3I(1+x)}
\right]^{1/2},
\ee
\be
\label{nc_2}
N_{c,{\rm eff}}'={N_c'\over\sqrt{1+x^{-1}}}=m'
\left[{B_{\alpha'}\mu a^2\over 3I'(1+x^{-1})}\right]^{1/2}.
\ee
Obviously, if the two stars have identical resonances ($x=1$), then 
$N_{c,{\rm eff}}$ would be smaller than $N_c$ by a factor of $\sqrt{2}$. If the energy dissipation rates in the two stars differ by
at least a factor of a few (e.g., $x\sim 0.2$), then
$N_{c,{\rm eff}}$ is only slightly modified from $N_c$, while 
$N_{c,{\rm eff}}'$ will be a factor of a few smaller than $N_c'$.

\subsection{Numerical Examples of Evolution Toward Resonance}

For a given set of modes, the solution of the system
of equations (\ref{doms})-(\ref{de}) depends on the dimensionless parameters
$t_{d\alpha}/t_{sd}$, $N_c$ and $\gamma_\alpha P$, as well as the initial
conditions. In general, these parameters change
as the system evolves.

\subsubsection{A Simple Example}

\begin{figure*}
\vskip -2.5cm
\begin{center}
\includegraphics[scale=0.75]{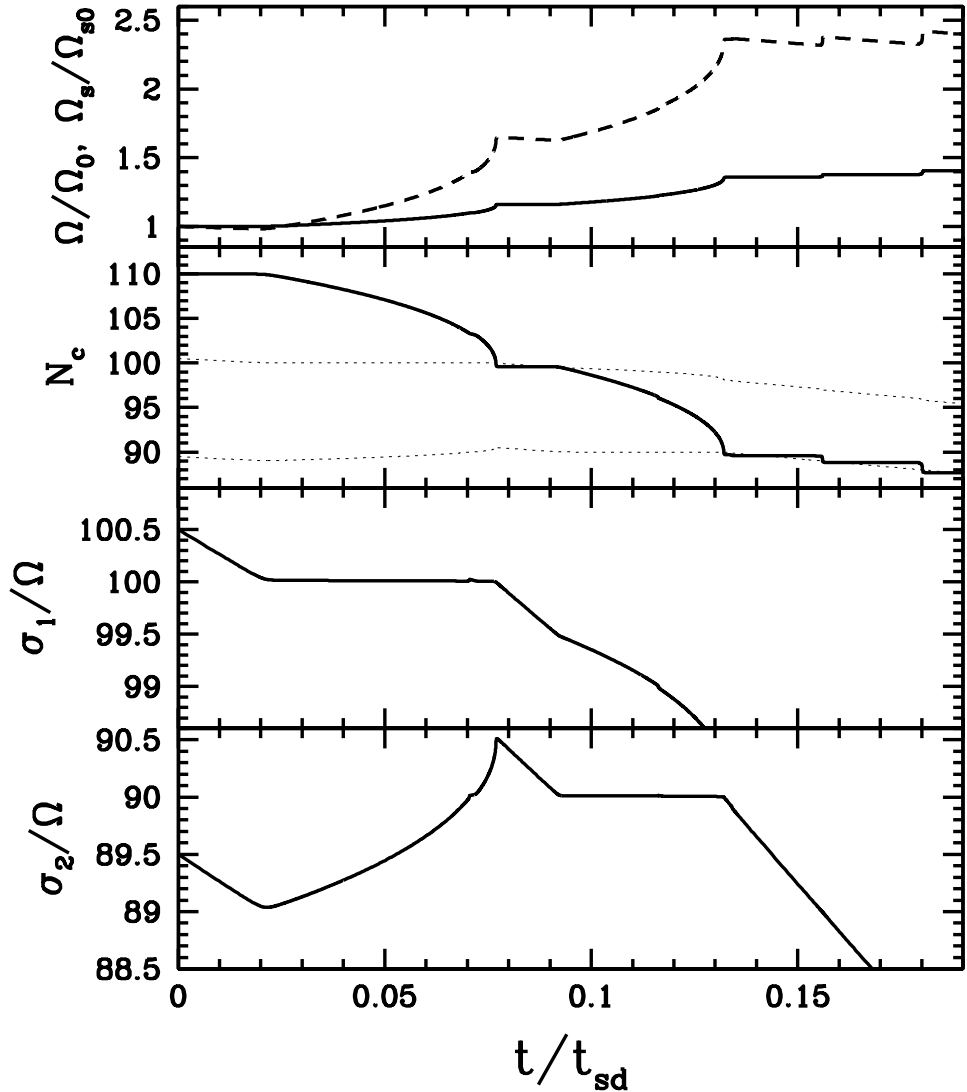}
\end{center}
\vskip -2.5cm
\caption{ \label{evolve} Secular evolution of the stellar spin and binary orbit 
driven by dynamical tides and ``intrinsic'' stellar spin-down.
Two modes are included, both having
$m=2$, $t_{d\alpha}/t_{sd}=100$ and $\gamma_\alpha P=0.01$.
The initial spin frequency is $\Omega_s=15\Omega$. The top panel
shows the orbital frequency (solid line) and the spin frequency 
(dashed line), both in units of their initial values. The second panel
shows $N_c$ as defined by equation (\ref{nc}). The bottom two panels show
the mode frequencies in units of $\Omega$. The two light dotted lines
on the third panel also show $\sigma_1/\Omega$ and $\sigma_2/\Omega$.
The time (on the horizontal axis) is expressed
in units of $t_{sd}$, the intrinsic spindown time of the star.
The system first evolves into resonance locking at
$\sigma_1=100\Omega$ and then a different resonance locking 
at $\sigma_2=90\Omega$. Note that resonance locking cannot occur when
$\sigma_\alpha/\Omega>N_c$.}
\end{figure*}

To illustrate the essential behavior of the secular evolution toward
resonance locking, we first consider the simple case where 
$t_{d\alpha}/t_d$ and $\gamma_\alpha P$ are assumed to be
constant and identical for all modes considered. Figure \ref{evolve} depicts
an example: we include two $m=2$ modes ($\sigma_1$ and $\sigma_2$), 
both have $t_{d\alpha}/t_{sd}=10^2$, $\gamma_\alpha P=10^{-2}$
and $B_\alpha=0.8$. The parameter $N_c$ has an initial value
of $110$, but we allow $N_c$ to evolve via $N_c\propto \Omega^{-2/3}$.
At $t=0$, the spin frequency is $\Omega_s=15\Omega$, and the two modes
have $\sigma_1/\Omega=100.5$ and $\sigma_2/\Omega=89.5$ (i.e., both are
initially ``off-resonance''). We see that the stellar spindown first 
causes $\sigma_1$ to lock into resonance at $\sigma_1/\Omega=100$.
The star then spins up, driven by the resonant tidal torque. The second
mode passes through the $N=90$ resonance, but it produces a negligible
effect on the $\sigma_1/\Omega=100$ resonance. In the 
meantime, the orbit decays, reducing $N_c$. When $N_c$ becomes
less than $100$, the resonance locking can no longer be maintained,
and $\sigma_1$ breaks away from the $N=100$ resonance. The stellar
spin then decreases (due to the $1/t_{sd}$ term), which leads to 
$\sigma_2$ to capture into the $N=90$ resonance. Eventually,
as $N_c$ decreases (due the resonant tidal torque) below $90$,
the second mode breaks away from the $N=90$ resonance. This example
corroborates our analytical result of section 5.3, and demonstrates
that resonance locking can indeed be achieved and maintained for an extended
period during the evolution of the binary system.

\subsubsection{More Realistic Examples}
\label{realistic}

Having demonstrated how resonance locking can occur in a simplified system consisting of only two oscillation modes, we now investigate how resonance locking is likely to occur in the actual KOI-54 system. We solve the orbital evolution equations (\ref{doms})-(\ref{dsig}) using the actual values of $\varepsilon_\alpha$, $Q_\alpha$, and $\gamma_\alpha$ found in section \ref{modes} to calculate each $t_{d\alpha}$. We include eighteen modes in our equations: they are the $l=2$, $m=0$, $m = \pm 2$ modes for six values of $\varepsilon_\alpha$ such that the initial frequencies of the $m=2$ modes range from $40 \Omega$ to $170 \Omega$. We allow the value of $K_{lm}$ to evolve as the values of $\sigma_\alpha$, $\Omega$, and $e$ change with time. 

\begin{figure*}
\begin{center}
\includegraphics[scale=0.52]{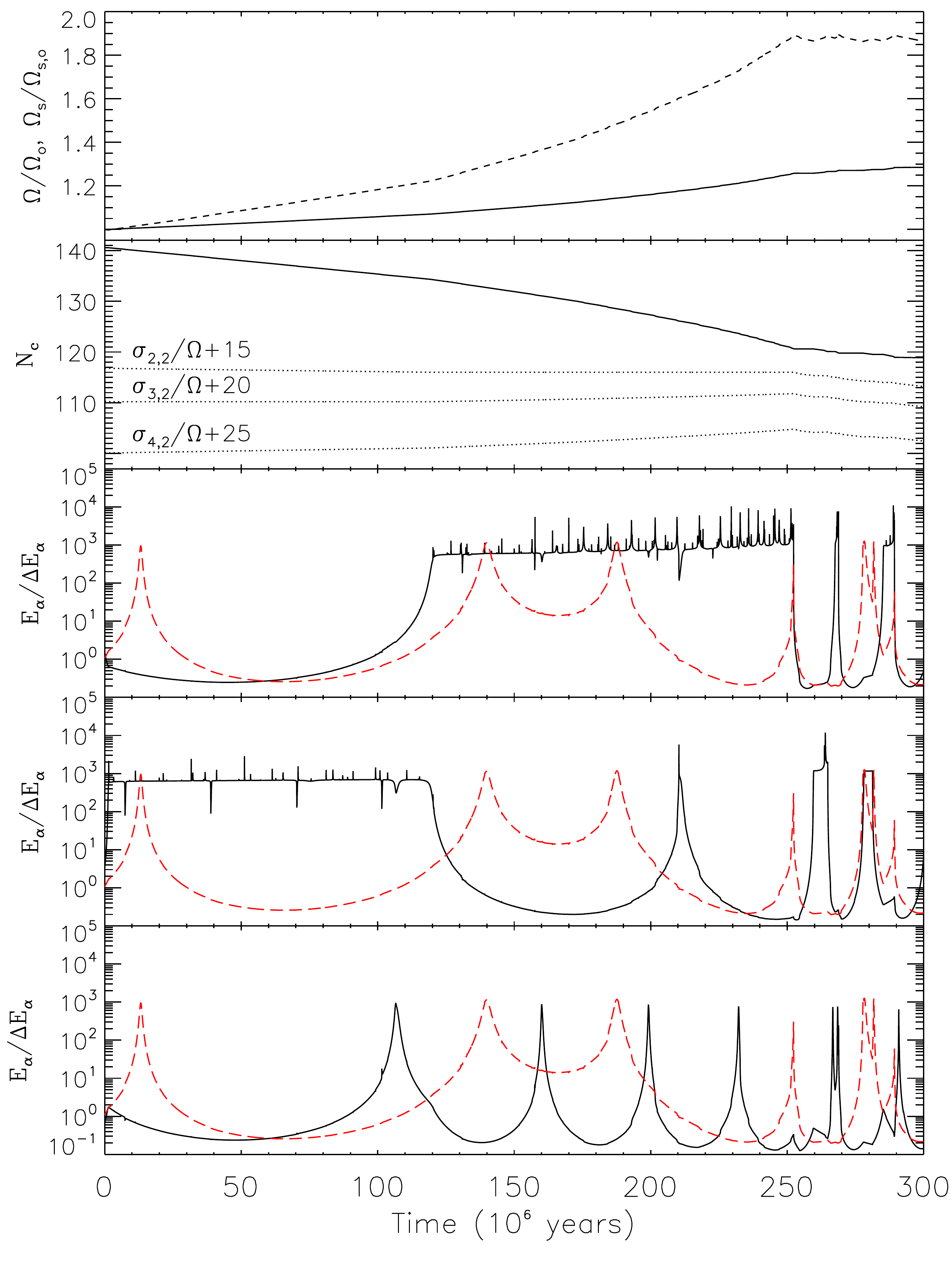}
\end{center}
\caption{\label{evolution} Secular evolution of the KOI-54 system driven by a sample of 18 oscillation modes and intrinsic stellar spindown.  Top: orbital frequency $\Omega$ (solid line) and spin frequency $\Omega_s$ (dashed line) in units of their initial values. Top middle: the value of $N_c$ (solid line) and the mode frequencies $\sigma_{2,2}/\Omega+15$, $\sigma_{3,2}/\Omega+20$, and $\sigma_{4,2}/\Omega+35$ (dotted lines) of the modes depicted in the bottom three panels. Middle: the mode energy $E_\alpha$ in units of $\Delta E_\alpha$ for the $\sigma_{2,2}$ mode (solid line) and the $\sigma_{4,0}$ mode (red dashed line). Bottom middle: same as middle except the solid line is for the $\sigma_{3,2}$ mode. Bottom: same as bottom middle except the solid line is for the $\sigma_{4,2}$ mode. The initial spin frequency is $\Omega_s \simeq 16.5 \Omega$, and the spindown time is $t_{sd} = 3\times10^8$ years.}
\end{figure*}

\begin{figure*}
\begin{center}
\includegraphics[scale=0.52]{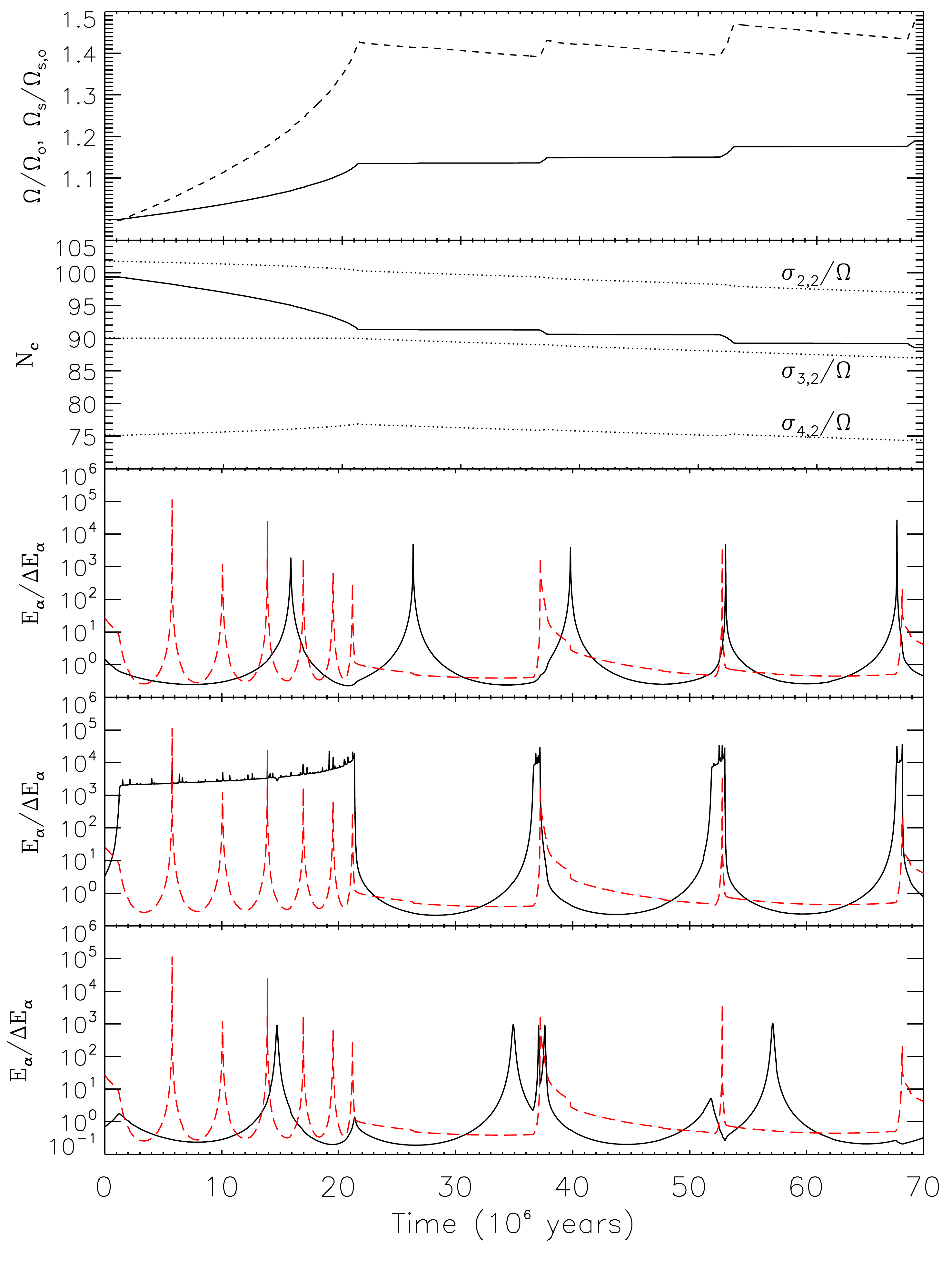}
\end{center}
\caption{\label{evolution2} Same as Figure \ref{evolution}, except this evolution doubles the orbital decay rate due to each mode, and the red dashed line corresponds to the $\sigma_{2,0}$ mode.}
\end{figure*}

Figures \ref{evolution} and \ref{evolution2} display the evolution of $\Omega$, $\Omega_s$, $N_c$, and the values of $\sigma_\alpha$ and $E_\alpha$ for a selected sample of modes over a time span of tens of millions years. We use initial values of $\Omega_o =0.95 \Omega_{\rm obs}$ (here $\Omega_{\rm obs}$ is the observed orbital frequency of KOI-54), $e_o=0.84$, $\Omega_{s,o} = \Omega_{ps} \simeq 16.5 \Omega_o$, and a spindown time of $t_{sd} = 3\times10^8$ years. In Figure \ref{evolution2}, we have doubled the orbital decay rate to account for an equal amount of energy dissipation in the companion star.

Let us start by examining Figure \ref{evolution}. The system quickly enters a resonance locking state with the $\sigma_{3,2}$ mode, where the notation $\sigma_{k,m}$ identifies the mode with the $k$th largest frequency in our simulation with azimuthal number $m$. The resonance locking lasts for over 100 million years, until the $\sigma_{2,2}$ mode reaches resonance. The $\sigma_{2,2}$ mode then locks in resonance for over 100 million years. During resonance locking, the orbital frequency and spin frequency increase rapidly. The energies $E_\alpha$ of the resonant locked modes can exceed $700$ their non-resonant values $\Delta E_\alpha$, corresponding to an increase in visibility of over 25 times the visibility during a non-resonant state. 
When no mode is locked in resonance, the orbital evolution is relatively slow, with the dominant effect being the spindown of the star due to the $t_{sd}$ term.

For the initial conditions and orbital evolution depicted in Figure \ref{evolution}, only two of our 18 modes pass through resonant locking phases: two $m=2$ modes with initial frequencies of $\sigma_{2,2,o} \simeq 103$ and $\sigma_{3,2,o} \simeq 90$. The highest frequency $m=2$ mode included in our evolution does not lock in resonance because it has $N>N_c$ at all times.  The lowest frequency modes do not lock in resonance because they do not satisfy equation (\ref{saturation}), i.e., they become saturated before they can begin to resonantly lock. This occurs because the value of $Q_\alpha$ for these high-order modes is small, resulting in a small orbital decay rate $t_{d \alpha}^{-1}$ for these modes. 

Which mode locks into resonance is dependent on the initial conditions. However, as can be seen in Figure \ref{evolution}, the resonance with the $\sigma_{2,2}$ mode ends the resonance with the $\sigma_{3,2}$ mode because the $\sigma_{2,2}$ mode has a larger value of $E_{\alpha}$ and will thus dominate the orbital decay. Therefore, we can expect the system to evolve into a state in which the mode with the largest value of $E_\alpha$ (and with $N<N_c$) is locked in resonance. 

In Figure \ref{evolution}, the resonance locking with the $\sigma_{2,2}$ mode is ended by a resonance with the $\sigma_{4,0}$ mode. The $\sigma_{4,0}$ mode does not lock into resonance because an $m=0$ mode cannot change the stellar spin frequency. However, its resonance causes the value of $\dot{\Omega}/\Omega$ to exceed the value of $\dot{\sigma}_\alpha/\sigma_\alpha$ for the locked mode, so that the system sweeps through the resonance. In other words, the non-locked resonant mode temporarily decreases the value of $t_d$, decreasing the value of $t_D$ such that resonant saturation occurs for the locked mode, causing it to pass through resonance.  Also note that even though the $\sigma_{4,0}$ mode does not lock into resonance, it maintains a relatively large energy ($E_\alpha > 10 \Delta E_\alpha$) for a period of over 50 million years. This is indicative of the pseudo-resonance locking for $m=0$ modes described in Section \ref{m0}.

Let us now examine Figure \ref{evolution2}, in which the orbital decay rate has been doubled (i.e., we multiply $\dot{\Omega}$ by 2) to account for an equal tidal response in the companion star. The results are significantly different: due to the increased orbital decay rate, the initial effective value of $N_{c,{\rm eff}}$ has dropped to $\sim 100$ (see section 5.3.3) so that the $\sigma_{2,2}$ mode can no longer lock into resonance. Also, the maximum energy $E_\alpha/\Delta E_\alpha$ of the modes is larger in this scenario (because the modes must enter deeper into resonance to become locked), so the modes can create larger magnitude variations while being locked in resonance.

However, the resonance locking events are shorter because the modes are nearly saturated during resonance locking, allowing resonances with $m=0$ modes or higher frequency $m=2$ modes to easily disrupt the resonance locking state. In the bottom three panels of Figure \ref{evolution2}, we have plotted the energy $E_\alpha$ of the $\sigma_{2,0}$ mode. In Figure \ref{evolution2}, three of the resonant locking events for the $\sigma_{3,2}$ mode are ended due to a resonance with the $\sigma_{2,0}$ mode. We have examined the results carefully and found that all of the resonance locking events were ended due a resonance with an $m=0$ mode or an $m=2$ mode for which $N>N_c$. Although the resonance locking events are brief at the end of the simulation, the locking event with the $\sigma_{3,2}/\Omega = 90$ mode lasts for about 20 million years at the beginning of the simulation when the system's parameters most closely resemble those of KOI-54. We thus conclude it is likely to observe a system such as KOI-54 in a resonance locking state.

We note the secular evolution analysis presented above does not take into account the fluctuation of mode amplitudes due to the changing strength of the tidal potential in an elliptical orbit. Therefore, in addition to the results displayed above, we have also performed calculations of the full dynamic evolution of the stellar oscillations and binary orbit, including the back-reaction of the modes on the orbit. We find that resonance locking can indeed be maintained for extended periods of time, and that non-secular effects have little impact on the results discussed above.

\section{Oscillations at Non-Orbital Harmonics}
\label{nonlinear}

As discussed in section 4, in the linear theory, tidally-forced stellar oscillations give rise to flux variations at integer multiples of the orbital frequency. Many of these oscillations have been observed in the KOI-54 system.  However, Welsh et al. (2011) also reported the detection of nine modes that do not have frequencies that are integer multiples of the orbital period, and their observation requires an explanation.

One possibility is that one (or both) of the stars in KOI-54 are $\delta$-Scuti variable stars. The masses, ages, and metallicities of the KOI-54 stars put them directly in the instability strip and so $\delta$-Scuti-type pulsations are not unexpected. However, as pointed out by Welsh et al. (2011), the complete absence of any modes detected with periods less than 11 hours is troublesome for the $\delta$-Scuti interpretation, as most observed $\delta$-Scuti pulsations have periods on the order of a few hours. One would then have to explain why only long period ($P>11$ hours) $\delta$-Scuti-type oscillations are visible in this system.

Another possibility is that some (or all) of the non-harmonic modes are excited via nonlinear couplings with resonant modes. In particular, it is thought that three-mode resonant coupling (parametric resonance) plays an important role in limiting the saturation amplitudes of overstable g-modes in ZZ-Ceti stars (Wu \& Goldreich 2001) and $\delta$-Scuti stars (Dziembowski \& Krolikowska 1985). In the KOI-54 system, when a resonantly excited mode reaches sufficiently large amplitude, it will couple to two non-resonant, lower-frequency daughter modes (see Kumar \& Goodman 1996), thereby explaining the observed non-harmonic oscillations. 

Parametric resonance can occur when $\omega_\gamma \simeq \omega_\alpha+\omega_\beta$, where $\omega_\gamma$ is the frequency of the parent mode and $\omega_\alpha$ and $\omega_\beta$ are the frequencies of the two daughter modes. The additional requirement that $m_\gamma = m_\alpha +m_\beta$, where $m$ is the azimuthal number of the mode, implies that $\sigma_\gamma \simeq \sigma_\alpha+\sigma_\beta$. Examination of the KOI-54 data given in Welsh et al. (2011) reveals that $f_2 = f_5+f_6$ (25.195 $\mu$Hz = 6.207 $\mu$Hz + 18.988 $\mu$Hz) exactly to the precision of the measurements, where $f_p$ is the frequency of the magnitude oscillation with the $p$th largest magnitude variation. This provides strong evidence that the $p=5$ and $p=6$ oscillations are due to modes excited via parametric resonance and not via direct tidal forcing.

However, no other pair of observed non-harmonic magnitude oscillations have frequencies which add up to that of an observed oscillation. Some of these non-resonant oscillations may be due to modes excited via parametric resonance that have undetected sister modes. This possibility is especially appealing if one considers the scenario in which one of the two daughter modes has $m=0$ (and could thus be easily detected for small values of $i_s$), while its sibling has $m=2$ (and is thus very difficult to detect for small $i_s$). We therefore suggest that all non-harmonic flux oscillations are produced by nonlinear mode coupling. Deeper observations (with better photometry) and analysis may reveal additional evidence for nonlinear mode couplings in the KOI-54 system

\section{Discussion}

We have shown that many properties of flux oscillations detected in
the KOI-54 binary system by {\it KEPLER} can be understood using the
theory of dynamical tidal excitations of g-modes developed in this
paper. In particular, our analysis and calculation of the resonance
mode locking process, which is driven by dynamical tides and intrinsic
stellar spindown, provides a natural explanation for the fact that
only those modes with frequencies ($\sigma_\alpha$) less than about
90-100 times the orbital frequency ($\Omega$) are observed. Our result
suggests that the KOI-54 system is currently in a resonance-locking
state in which one of the stars has a rotation rate such that it
possesses a $m=2$ mode with frequency $\sigma_\alpha=90\Omega$ and the
other star has a similar mode with $\sigma_{\alpha'}=91\Omega$ ---
these modes produce the largest flux variations detected in
KOI-54. Our analysis shows that the system can evolve into and stay in
such a resonance locking state for relatively long time periods, and it is
reasonable to observe the system in such a state.  Other less prominent
flux variations at lower frequencies can be explained by the $m=0$ tidally
forced oscillations, many of which may be enhanced by the resonant
effect. We have also found evidence of nonlinear three-mode coupling
from the published {\it KEPLER} data and suggested that the nonlinear
effect may explain the flux variabilities at non-harmonic frequencies.

In our study, we have used approximate quasiadiabatic mode damping rates. Obviously, it
would be useful to repeat our analysis using more realistic mode damping
rates, calculated with the full non-adiabatic oscillation equations. Our calculations of the flux variations of tidally induced
oscillations have also highlighted the importance of accurate
treatment of non-adiabatic effects in the stellar photosphere. We
have only considered g-modes (modified by stellar rotation) in this
paper. Our general theory allows for other rotation-driven modes, such
as inertial/Rossby modes. It would be interesting to study tidal
excitations of these modes as well as the nonlinear 3-mode coupling 
effect in the future.

More detailed modelling of the observed oscillations may provide useful constraints
on the parameters of the KOI-54 system, particularly the stellar rotation rates 
and spin inclinations. On a more general level, KOI-54 may serve as a ``laboratory''
for calibrating theory of dynamical tides, which has wide applications
in stellar and planetary astrophysics (see references in Section 1).

While we believe that our current theory provides the basis for
understanding many aspects of the KOI-54 observations, some puzzles
remain.  In our current interpretation, the most prominent
oscillations (at $90\Omega$ and $91\Omega$) occur in the two different
stars, each having a $m=2$ mode resonantly locked with the
orbit. While one can certainly appeal to coincidence, it is intriguing
that the strongest observed oscillations are at the two consecutive
harmonics ($N=90,\,91$) of the orbital frequency. More importantly, as
discussed in Section 5.3.3, when both stars are involved in resonance
locking, the effective $N_c$ (above which resonance locking cannot
happen) is reduced. For example, using our canonical stellar
parameters ($\kappa =0.05$), we find $N_c\simeq 131$ [see equation (\ref{nc2})], but the effective
$N_c$ would be reduced to $92$ if the two stars have identical modes in
resonance locking. In the case where the resonant energy
dissipation rates in the two stars differ by a factor of more than a
few, the effective $N_c$ of one star would remain close to 131, while the
effective $N_c$ for the other star would likely be reduced to a value
below 90. In this case, it would be unlikely to find the binary
system in the resonance-locking state involving both stars with
$N,N'$ close to 90, because it is most likely for modes with $N$ just less than $N_c$ to be locked in resonance (see Section 5.4.2.). Thus, we find it likely that the energy dissipation rates are nearly equal (to within a factor of 2) in each star in the KOI-54 system.

Finally, we may use our results to speculate on the evolutionary history
(and future) of the KOI-54 system. The star with $M_2=2.38 M_\odot$
and radius $R_2 =2.33 R_\odot$ has an age about $3\times
10^8$ years, according to our stellar model generated by the MESA
code. During the resonance locking phase, the orbital
eccentricity decreases on a relatively short time scale (of order $\sim 10^8$ years). The current large eccentricity of KOI-54 then suggests that resonance locking has not operated for a large fraction of the system's history. When resonance locking is in effect, 
the orbital evolution time scale is set by primarily by the
spindown time scale, $t_{sd}$, and the value of $\delta_N$ for the
resonant mode [see equation (\ref{domres})].
In the future, the stars will continue to expand
into red giants and the spindown time scale will decrease. This will
cause the orbital evolution time scale to correspondingly decrease, leading to 
rapid orbital decay and circularization of the system.

While our paper was under review,  a paper by Burkart et al. (2011) appeared 
on arXiv.  Burkart et al. carried out non-adiabatic mode calculations and also attribute most of the oscillations to $m=0$ modes, although they do not reach a definite conclusion on the source of the $N=90,91$ oscillations. They did not consider resonance locking (instead they consider the qualitatively different phenomenon of \textquotedblleft pseudosynchronous locks"), and appeared to attribute all resonances to random chance. They also showed that 3-mode coupling (see Section 6) is possible.

\section*{Acknowledgments}
This work has been supported in part by the NSF grant AST-1008245.


\def\apj{{Astrophys. J.}}
\def\apjs{{Astrophys. J. Supp.}}
\def\mnras{{Mon. Not. R. Astr. Soc.}}
\def\prl{{Phys. Rev. Lett.}}
\def\prd{{Phys. Rev. D}}
\def\apjl{{Astrophys. J. Let.}}
\def\pasp{{Publ. Astr. Soc. Pacific}}
\def\aapr{{Astr. Astr. Rev.}}

\end{document}